\documentstyle[12pt]{article}
\makeatletter
\@addtoreset{equation}{section}
\makeatother

\renewcommand{\baselinestretch}{1.2}

\newcommand{\startappendix}{
\renewcommand{\thesection}{\Alph{section}}}

\newcommand{\wha}{\widehat{\cal A}}

\newcommand{\delb}{\bar{\partial}}

\newcommand{\bz}{\bar{z}}

\newcommand{\dis}{\displaystyle}

\newcommand{\be}{\begin{equation}}
\newcommand{\ee}{\end{equation}}
\newcommand{\bea}{\begin{eqnarray}}
\newcommand{\ena}{\end{eqnarray}}
\newcommand{\eea}{\end{eqnarray}}
\newcommand{\beano}{\begin{eqnarray*}}
\newcommand{\enano}{\end{eqnarray*}}
\newcommand{\sect}[1]{\setcounter{equation}{0}\section{#1}}
 \newcommand{\hs}[1]{\hspace{#1
mm}}
\newcommand{\half}{{\frac{1}{2}}}
\newcommand{\shalf}{\mbox{$\frac{1}{2}$}} 

\newcommand{\ca}{\mbox{$\cal{A}$}}

\newcommand{\cd}{\mbox{$\cal{D}$}}
\newcommand{{\cf}}{\mbox{$\cal{F}$}}
\newcommand{{\cg}}{\mbox{$\cal{G}$}}
\newcommand{\ch}{\mbox{$\cal{H}$}}

\newcommand{\cq}{\mbox{$\cal{Q}$}}
\newcommand{\cs}{\mbox{$\cal{S}$}}

\newcommand{\cv}{\mbox{$\cal{V}$}}

\newcommand{\ssa}{sub-superalgebra}
\newcommand{\ls}{\mbox{$s\ell(1|2)$}}
\newcommand{\osp}{\mbox{$osp(1|2)$}}
\newcommand{\prf}{\underline{Proof:}\ }

\newcommand{\pp}{{=\!\!\! |}}
\newcommand{\del}{\partial}
\newcommand{\bdel}{\bar{\partial}}

\newcommand{\und}[1]{\underline{#1}}

\newcommand{\mb}[1]{\hs{5}\mbox{#1}\hs{5}}

\newcommand{\su}{\mbox{${s\ell(2)}$}}
\newcommand{\qq}{{\cal Q}}
\newcommand{\rinde}{[(0,\half)]^2_{\bf A}}

\newtheorem{theor}{Theorem}

\newtheorem{lem}{Lemma}
\newcommand{\Z}{\mbox{$Z\hspace{-2mm}Z$}}
\newcommand{\N}{\mbox{\hspace{.04mm}\rule{0.2mm}{2.8mm}\hspace{-1.5mm} N}}
\newcommand{\Asym}[1]{\left[\rule{0mm}{5mm} #1\right]_{\bf A}}
\newcommand{\Sym}[1]{\left[\rule{0mm}{5mm} #1\right]_{\bf S}}
\newcommand{\r}[1]{\mbox{$({#1},{#1})$}}
\newcommand{\rge}[2]{{\mbox{(${#1}$,${#2}$)}}}
\newcommand{\rpi}[1]{\mbox{$({#1},{#1})^\pi$}}
\newcommand{\rmo}[1]{\mbox{(-${#1},{#1})$}}
\newcommand{\rpim}[1]{\mbox{(-${#1},{#1})^\pi$}}
\newcommand{\ro}[1]{\mbox{$(0,{#1})$}}
\newcommand{\ropi}[1]{\mbox{$(0,{#1})^\pi$}}
  
\newcommand{\CMP}[1]{Comm.\ Math.\ Phys.\ {\bf #1}}
\newcommand{\JMP}[1]{Journ.\
Math.\ Phys.\ {\bf #1}}

\def\vf{\varphi}
\def\a{\alpha}
\def\b{\beta}

\def\d{\delta}
\def\e{\epsilon}

\def\g{\gamma}
\def\h{\eta}

\def\j{\psi}
\def\k{\kappa}

\def\p{\pi}

\def\t{\tau}

\def\P{\Pi}

\newcommand{\nonu}{\nonumber \\[2mm]}

\begin{document}
%
%% FOLLOWING LINE CANNOT BE BROKEN BEFORE 70 CHAR
%------------------------------title-----------------------------------
\begin{titlepage}
\topskip0.5cm
\hfill\hbox{ENSLAPP-555/95}\\[-1.4cm]
\flushright{\hfill\hbox{VUB-TH.395}}\\
\flushright{\hfill\hbox{hep-th/9511049}}\\[2.3cm] \begin{center}{\Large\bf
Strings from $N=2$ Gauged Wess-Zumino-Witten Models\\[1.5cm]}{ \Large E.
Ragoucy${}^1$, A. Sevrin${}^2$ and P.
Sorba${}^1$\\[.6cm]}
${}^1$ Laboratoire de Physique, ENSLAPP, Annecy, France\\
${}^2$ Theoretische Natuurkunde, Vrije Universiteit Brussel, B-1050 Brussels,
Belgium.\\
\end{center}
\begin{abstract}
We present an algebraic approach to string theory.  An embedding of $sl(2|1)$
in a super Lie algebra together with a grading on the Lie algebra determines a
nilpotent subalgebra of the super Lie algebra. Chirally gauging this subalgebra
in the corresponding Wess-Zumino-Witten model, breaks the affine symmetry of
the Wess-Zumino-Witten model to some extension of the $N=2$ superconformal
algebra. The extension is completely determined by the $sl(2|1)$ embedding. The
realization of the superconformal algebra is determined by the grading. For a
particular choice of grading, one obtains in this way, after twisting, the BRST
structure of a string theory. We classify all embeddings of $sl(2|1)$ into Lie
super algebras and give a detailed account of the branching of the adjoint
representation. This provides an exhaustive classification and characterization
of both all extended $N=2$ superconformal algebras and all string theories
which can be obtained in this way.
\end{abstract}
\vfill
%\hbox{CERN-TH.xxxxxxx}\hfill\\
\hbox{November 1995}\hfill\\
\end{titlepage}
%\end{document}
%
% ------------------- ---------------------------------------- %
\renewcommand{\baselinestretch}{1.2}\large\normalsize
\section{Introduction}
By now, there is a pleithora of different string theories. One way to
categorize
them is according to
the gauge algebra on the worldsheet. Taking the Virasoro algebra as gauge
algebra, one obtains the
bosonic string, the N=1 super Virasoro algebra gives rise to the superstring,
the $W_n$ algebra yields
$W_n$-strings, ... In fact, it might that to each extension of the
Virasoro algebra, one can associate
a string theory\footnote{See however \cite{KAP} for a potential counter
example.}.
Given a gauge algebra, there is still a very large freedom which
consists of the particular
choice of the realization or string vacuum. Though all of these string theories
are perfectly consistent in
perturbation theory, only a very restricted set gives rise to
phenomenologically
acceptable theories.
However, as long as we do not understand the non-perturbative behaviour of
string theory, one should
study all classical solutions, hoping that this provides hints to the real
structure of string theory.
Some glimpse of a more systematic structure was seen in \cite{BV}, where it was
shown that the bosonic
string is a special choice of vacuum of an $N=1$ superstring, the $N=1$
superstring is then a special choice
of vacuum of the $N=2$ superstring. Similar patterns, involving other types of
string theories were obtained
later on. Though quite a fascinating observation, its relevance remains to be
understood (see {\it e.g.} the
remarks in \cite{pol}).
A seemingly unrelated approach was initialized in \cite{bea}. There it was
shown
that the BRST structure
of the bosonic string is encoded in a twisted $N=2$ superconformal algebra.
This
seems to be a universal
feature of string theories: for any string theory, the BRST structure is given
by a twisted extension of the
$N=2$ superconformal algebra. This has been worked out in several concrete
cases: the BRST strucure of
the superstring is given by the $N=3$ superconformal algebra \cite{BLNW}, the
$W_n$ strings by the
corresponding twisted $N=2$ $W_n$ algebra \cite{BLNW} and strings with $N$
supersymmetries have
the Knizhnik-Bershadsky $SO(N+2)$ superconformal symmetry \cite{BLLS}. Even
topological strings exhibit such a
structure \cite{LLS}
The main idea here is that one adds both the BRST current and the anti-ghosts
to
the gauge algebra.
The BRST-charge itself is then one of the supercharges,
\bea
{{\cal Q}_{BRST}}\ \equiv\ G^+_0\ =\ \frac{1}{2\pi i}\oint dz\,(cT+\dots)\ ,
\eea
while the Virasoro anti-ghost $b(z)$ is the conjugate supercurrent $G^-(z)$.
This automatically ensures
that $T(z)=\{{{\cal Q}_{BRST}},b(z)\}$. Together this generates the twisted
$N=2$ superconformal
algebra. For a string theory with a larger gauge algebra, one gets in this way
some twisted extension of
the $N=2$ superconformal algebra.
Once, the presence of this twisted $N=2$ structure is accepted, one might try
to
use it to {\it define} the
string theory. This gives rise to a very algebraic and systematic approach
to string theory. The
obvious way to achieve such a systematics is through gauging Wess-Zumino-Witten
models, which is also known as
quantum Hamiltonian reduction. It is well
known that the reduction
of $sl(2|1)$ gives rise to the $N=2$ superconformal algebra. Embeddings of
$sl(2|1)$ in super Lie algebras
yield then all extensions of the $N=2$ algebra which can be obtained through
Hamiltonian reduction.
Once a particular embedding is given, the extended $N=2$ superconformal algebra
is uniquely determined.
However, the particular (free field) realization (or quantum Miura transform)
one obtains for this algebra
is determined by a choice of a grading on the super Lie algebra. As we will
see,
only a very particular
grading allows for a stringy interpretation.
This approach to string theory has the great advantage that it is almost
completely algebraic. The
calculations are of an algorithmic nature, enabling one to obtain {\it e.g.}
the
explicit form of the
BRST current in a straightforward way (compare this to the usual trial and
error method). This
program has been explicitely carried out for the non-critical $W_n$ strings,
based on a reduction of
$sl(n|n-1)$ \cite{BLNW} and strings with $N$ supersymmetries, based on a
reduction of $osp(n+2|2)$
\cite{BLLS}. In the former case, only classical arguments were given, while in
the latter, at least for $N=1$ and 2, the full quantum
structure was exhibited.
As $N=2$ superconformally invariant models exhibit a very rich structure, (for
an extensive review, see \cite{fre}), the full classification of extended $N=2$
algebras obtainable from Hamiltonian reduction is in itself an interesting
result.
In the present paper we start in the next section with the simplest example
available: the bosonic string.
We show that the BRST structure is indeed given by a twisted $N=2$ algebra and
we derive it from
the reduction of $sl(2|1)$. This example exhibits already many of the
complications which arise in the
general case. Section 3 classifies all $sl(2|1)$ embeddings in super Lie
algebras. In section 4, a detailed
study is made of the branching of the adjoint representation of a super Lie
algebras into irreducible
representations of the embedded $sl(2|1)$ algebra. This basically
determines the field
content of the extended $N=2$ superconformal algebra. The results of section 4
are
applied in the next sections. In section 5 we briefly discuss the reduction
using the standard grading. This yields the so-called
symmetric realizations of the extended
$N=2$ superconformal algebra. In section 6 we determine,
given some embedding of $sl(2|1)$ in a super Lie algebra, the grading which
will
yield a ``stringy'' reduction. We construct the gauged WZW model
which describes the reduction. In the next section the model is quantized and
the resulting
string theory is discussed. We end with some
conclusions and open problems. In the appendix we summarize some properties of
WZW models.
\section{A Simple Example}
Before treating the general case, we illustrate the main ideas with the
simplest
example: the bosonic string.
We give this example in considerable detail as it exhibits many of the
complications which arise in the
general case. The string consists of a matter, a gravity or Liouville and a
ghost sector. At this point, we
do not make any assumptions about the particular structure of the matter
sector.
We just represent it by
its energy-momentum tensor $T_m$ which generates the Virasoro algebra with
central charge $c_m$.
The gravity sector is realized in terms of a Liouville field $\vf_L$,
$\del\vf_L(z_1)\del\vf_L(z_2)=-z_{12}^{-2}$,
with energy-momentum tensor $T_L$:
\bea
T_L=-\frac 1 2 \del \vf_L\del
\vf_L+\sqrt{\frac{25-c_m}{12}}\del^2\vf_L, \eea
which has central charge $c_L=26-c_m$. The energy-momentum tensor for the ghost
system assumes the
standard form:
\bea
T_{gh}=-2B\del C-(\del B) C,
\eea
and has central extension $c_{gh}=-26$. The total energy-momentum tensor $T =
T_m + T_L + T_{gh}$
has central charge 0. The BRST current
\bea
J_{BRST}=C\left( T_m+T_L+\frac 1 2 T_{gh}\right)+\a \del \left(C\del
\vf_L\right) + \b \del^2 C,
\label{JBRS}
\eea
with
\bea
\a&=&-\frac{\sqrt{3}}{6}\left(\sqrt{1-c_m}+\sqrt{25-c_m}\right) \nonu
\b&=&-\frac{1}{12}\left( 7-c_m +\sqrt{(1-c_m)(25-c_m)}\right)
\eea
has only regular terms in its OPE with itself:
$J_{BRST}(z_1)J_{BRST}(z_2)=\cdots$. Note that this is a stronger statement
than
nilpotency
of the BRST charge $\qq_{BRST}^2=0$ with
\bea
\qq_{BRST}=\frac{1}{2\pi i}\oint dz J_{BRST}. \eea
The total derivative terms in Eq. (\ref{JBRS}), wich have no influence on the
BRST operator, have precisely been added to
achieve this \cite{bea}. Calling
$G_+\equiv J_{BRST}$ and $G_-\equiv B$, one finds that the current algebra
generated by $T$, $G_+$
and $G_-$ closes, provided a $U(1)$ current $U$ is introduced: \bea
&&T(z_1)T(z_2)=2z_{12}^{-2}T(z_2)+z_{12}^{-1}\del T(z_2),\qquad
T(z_1)G_+(z_2)=z_{12}^{-2}G_+(z_2)+z_{12}^{-1}\del G_+(z_2),\nonu
&&T(z_1)G_-(z_2)=2z_{12}^{-2}G_-(z_2)+z_{12}^{-1}\del G_-(z_2),\nonu
&&T(z_1)U(z_2)=-\frac{c_{N=2}}{3} z_{12}^{-3} + z_{12}^{-2} U(z_2)+
z_{12}^{-1}\del U(z_2),\nonu
&&G_+(z_1)G_-(z_2)= \frac{c_{N=2}}{3}z_{12}^{-3}+z_{12}^{-2}U(z_2) +
z_{12}^{-1} T(z_2),\nonu
&&U(z_1)G_\pm (z_2)=\pm z_{12}^{-1}G_\pm (z_2), \qquad
U(z_1)U(z_2)=\frac{c_{N=2}}{3}z_{12}^{-2}, \label{twn2}\eea
where $U$ is a modification of the ghost number current:
\bea
U\equiv -BC-\a\del\vf_L,
\eea
and
\bea
c_{N=2}=6\b .
\eea
Upon untwisting $T_{N=2}=T-\frac 1 2 \del U$, one gets the standard $N=2$
superconformal
algebra with central extension $c_{N=2}=6\b$.

For the critical bosonic string, $c_m=25$, we get
in this way $c_{N=2}=9$. Taking the Virasoro minimal models for the matter
sector:
\bea
c_m=1-6\left(\sqrt{\frac p q}- \sqrt{\frac q p}\right)^2, \eea
one gets
\bea
c_{N=2}=3\left( 1 - 2 \frac p q \right). \eea
In particular, taking $(p,q) = (1,k+2 )$, we get $c_{N=2}=3k/(k+2)$, {\it i.e.}
the $N=2$ minimal models, a fact which was heavily used in \cite{BLNW}.

We now turn to the Hamiltonian reduction and show how to obtain the above from
it. The super Lie
algebra $sl(2|1)$ is generated by a bosonic $su(2)\oplus u(1)$ sector: $\{
e_\pp\, ,e_=,e_0,u_0\}$
with $[e_0,e_\pp\, ]=+2e_\pp$, $[e_0,e_=]=-2e_=$ and $[e_\pp\, ,e_=]=e_0$. The
fermionic generators,
$g_\pm$, $\bar{g}_\pm$ are $sl(2)$ doublets, while $g$ ($\bar{g}$) has
eigenvalue $+1$ ($-1$) under
$\mbox{ad}_{u_0}$. The remaining commutation relations are easily derived from
the $3\times 3$ matrix
representation\footnote{$E_{kl}$ is a matrix unit, {\it i.e.} $(E_{kl})_{rs}=
\d_{kr}\d_{ls}$.}
$e_\pp=E_{12}$, $e_==E_{21}$, $e_0=E_{11}-E_{22}$,
$u_0=-E_{11}-E_{22}-2E_{33}$,
$g_+=E_{13}$, $g_-=E_{23}$, $\bar{g}_+=E_{32}$ and $\bar{g}_-=E_{31}$.
The WZW model, with action $\k S^- [g]$ on $sl(2|1)$ gives rise to affine
currents
$J=E^\pp e_\pp + E^0 e_0 + E^=e_= + U^0 u_0 + F^+ g_+ +F^- g_- +\bar{F}^+
\bar{g}_++\bar{F}^- \bar{g}_-$ which satisfy the OPE's\footnote{We use the
metric
$g_{ab}=-2str(t_a t_b)$.}
\bea
E^0 (z_1)\,E^0(z_2) = \frac{\kappa}{8}z_{12}^{-2}\, ,&\quad&
U^0 (z_1)\,U^0(z_2) = -\frac{\kappa}{8}z_{12}^{-2}\,,\nonu
E^0 (z_1)\,E^\pp (z_2) = \frac 1 2 z_{12}^{-1}E^\pp (z_2)\, , &\quad&
E^0 (z_1)\,E^= (z_2) = -\frac 1 2 z_{12}^{-1}E^= (z_2)\, ,\nonu
E^\pp (z_1)\,E^=(z_2) = \frac{\kappa}{4}\,z_{12}^{-2} + z_{12}^{-1}E^0 (z_2)\,
,\nonu
E^0(z_1)F^{\pm }(z_2) = \pm \frac{1}{4}z_{12}^{-1} F^{\pm }(z_2)\, , &\quad&
E^0(z_1)\bar{F}^{\pm }(z_2) = \pm \frac{1}{4}z_{12}^{-1} \bar{F}^{\pm }(z_2)\,
,\nonu
U^0(z_1)F^{\pm }(z_2) = - \frac{1}{4}z_{12}^{-1} F^{\pm }(z_2)\, , &\quad&
U^0(z_1)\bar{F}^{\pm }(z_2) = + \frac{1}{4}z_{12}^{-1} \bar{F}^{\pm }(z_2)\,
,\nonu
E^\pp(z_1)F^{-}(z_2) =+ \frac{1}{2} z_{12}^{-1}F^{+}(z_2)\, ,&\quad&
E^=(z_1)F^{+}(z_2) =+ \frac{1}{2} z_{12}^{-1}F^{-}(z_2)\, ,\nonu
E^\pp(z_1)\bar{F}^{-}(z_2) =- \frac{1}{2} z_{12}^{-1}\bar{F}^{+}(z_2)\,
,&\quad&
E^=(z_1)\bar{F}^{+}(z_2) =- \frac{1}{2} z_{12}^{-1}\bar{F}^{-}(z_2)\, ,\nonu
F^+(z_1)\bar{F}^+(z_2)=\frac 1 2 z_{12}^{-1}E^\pp(z_2)\, , &\quad&
F^-(z_1)\bar{F}^-(z_2)=\frac 1 2 z_{12}^{-1}E^=(z_2)\, ,\nonumber \eea
\bea
F^+(z_1)\bar{F}^-(z_2)=\frac \k 4 z_{12}^{-2}+\frac 1 2 z_{12}^{-1} E^0(z_2)+
\frac 1 2 z_{12}^{-1}U^0(z_2)\, ,\nonu F^-(z_1)\bar{F}^+(z_2)=+\frac \k 4
z_{12}^{-2}-
\frac 1 2 z_{12}^{-1} E^0(z_2)+\frac 1 2 z_{12}^{-1}U^0(z_2)\, .
\label{sl2cc}\eea

To perform the Hamiltonian reduction one has to introduce a grading and
constrain the strictly
negatively graded part of the current. This will give rise to a gauge symmetry
generated by the
strictly positively graded subalgebra of the super Lie algebra. Usually one
takes the grading to
be the one given by $\frac 1 2\mbox{ad}_{e_0}$ \cite{BTVD}, which, in our case
gives,\\[.1cm]
\begin{center}
{ \begin{tabular}{|c||c|c|c||c||c|c||c|c||}\hline $
$&$E^\pp$&$E^0$&$E^=$&$U^0$&$F^{+}$&
$F^{-}$&$\bar{F}^{+}$&$ \bar{F}^{-}$ \\ \hline
${\it grade}$&$1$&$0$&$-1$&$0$&$1/2$&$-1/2$&$1/2$&$-1/2$\\ \hline
\end{tabular}}\ .\\[1cm]
\end{center}
The constraint one imposes on the affine current is then simply $\Pi_{<0} J =
\frac \k 2(e_=+\t g_-+\bar{\t}\bar{g}_-)$, where $\Pi_{<0}$ projects on the
strictly negatively
graded part of the Lie algebra and $\t$ and $\bar{\t}$ are auxiliary fields
needed to obtain first class
constraints\footnote{We use a slightly confusing notation in eq. (\ref{ddeff}),
as $\t$ denotes both
a Lie algebra valued field and one of its components. The context should make
it clear what is meant by $\t$.}. These constraints follow from the action \bea
{\cal S}=\k S^-[g]+\frac{1}{\pi x}\int str\, A(J-\frac \k 2 e_=-\frac \k 2
[e_=,{\tau}])
-\frac{\k}{4\p x}\int str [e_=,{\tau}] \bdel{\tau},
\eea
where
\bea
A&=&A^\pp e_\pp+A^+g_++\bar{A}^+\bar{g}_+,\nonu {\tau}&=&\t
g_++\bar{\t}\bar{g}_+.\label{ddeff}
\eea
The action has a gauge invariance parametrized by $h=\exp \h$, $\h\in\P_{>0}
sl(2|1)$ or $\h=\h^\pp
e_\pp+\h^+g_++\bar{\h}^+\bar{g}_+$,
\bea
g&\rightarrow& g'=hg,\nonu
A&\rightarrow& A'=\bdel h h^{-1} + h A h^{-1},\nonu {\tau}&\rightarrow&
{\tau}'={\tau} - \P_{\frac 1 2 }\h. \eea
Fixing this symmetry by putting $A=0$, we get, upon introducing ghosts
$c = c^\pp e_\pp+\g^+g_++\bar{\g}^+\bar{g}_+ \in \P_{>0} sl(2|1)$ and
anti-ghosts
$b = b^= e_= + \b^- g_- +\bar{\b}^- \bar{g}_- \in \P_{<0} sl(2|1)$, the gauge
fixed action:
\bea
{\cal S}_{gf}=\k S^- [g] -\frac{\k}{4\p x}\int\, str [e_=,{\tau}] \bdel
{\tau}
+\frac{1}{2\p x}\int str b\bdel c, \eea
and the BRST charge
\bea
Q_{HR}=\frac{1}{4\p i x}\oint str \left\{ c\left( J-\frac \k 2 e_= -\frac \k 2
[e_=,{\tau} ]
+ \frac 1 2 J_{gh} \right)\right\}, \eea
where $J_{gh}=1/2 \{ b, c\}$.
Because of the constraints, the original $sl(2|1)$ affine symmetry of the WZW
model breaks
down to an
$N=2$ superconformal symmetry. The generators of the $N=2$ superconformal
algebra are precisely
the generators of the cohomology ${\cal H}^*({\cal A},Q_{HR})$, where $\ca$ is
the algebra
generated by $\{ b, \hat{J}=J+ J_{gh}, {\tau}, c\}$ and all normal ordered
products of these fields
and their derivatives. In \cite{ST}, the computation of this cohomology has
been
done in general.
Applying the results and methods from \cite{ST}, one obtains \bea
T_{N=2}&=& \frac{2\k}{\k+1}\left(\hat{E}^\pp+
\hat{\bar{F}}^+\t+\hat{F}^+\bar{\t}
-\frac 2 \k \hat{U}^0 \hat{U}^0 +\frac 2 \k \hat{E}^0 \hat{E}^0\right.\nonu &&
\left. -\frac{\k+1}{\k}\del \hat{E}^0-\frac{\k+1}{4} \left( \t\del\bar{\t}
-\del\t \bar{\t}\right)\right)\nonu
G_+&=& \sqrt{\frac{4\k}{\k+1}}\left(\hat{F}^+-\t
\left(\hat{E}^0+\hat{U}^0\right)
+\frac{\k+1}{2}\del\t\right) \nonu
G_-&=&\sqrt{\frac{4\k}{\k+1}}\left(\hat{\bar{F}}^+
-\bar{\t }\left(\hat{E}^0 -\hat{U}^0\right)+\frac{\k+1}{2}\del\bar{\t}\right)
\nonu
U&=&-4\left( \hat{U}^0-\frac \k 4 \t\bar{\t}\right) , \eea
which satisfies the $N=2$ superconformal algebra with $c_{N=2}= -3(1+2\k)$.
The algebra $\ca$ has a natural double grading, \bea
\ca=\bigoplus_{\stackrel{\scriptstyle m,n\in\frac 1 2{\bf Z}}{m+n\in{\bf
Z}}}\ca_{(m,n)},
\eea
where in $(m,n)$, $m$ is the canonical grading used in the reduction and $m+n$
is the ghost number.
The auxiliary fields ${\tau}$ are assigned grading $(0,0)$. In \cite{ST} it
was proven that the map
$X\rightarrow \P_{(0,0)}X$, where $X=T_{N=2}$, $G_\pm$ or $U$, is an algebra
isomorphism.
This is the so-called Miura transform. Performing this map, we get the standard
free field realization of the $N=2$ superconformal algebra:
\bea
T_{N=2}&=& \del\vf\del\bar{\vf}-\frac{\sqrt{\k+1}}{2}\del^2(\vf+\bar{\vf})
-\frac 1 2 (\j\del\bar{\j}-\del\j \bar{\j})\nonu
G_+&=&- \j\del\vf+\sqrt{\k+1}\del\j\nonu
G_-&=&- \bar{\j}\del\vf+\sqrt{\k+1}\del\bar{\j} \nonu
U&=&\j\bar{\j}-\sqrt{\k+1}(\del\vf-\del\bar{\vf}),
\eea
where $\del\vf(z_1)\del\bar{\vf}(z_2)=z_{12}^{-2}$,
$\j(z_1)\bar{\j}(z_2)=z_{12}^{-1}$ and we
introduced some simple rescalings:
\bea
\del\vf=\frac{2}{\sqrt{\k+1}}(\hat{E}^0+\hat{U}^0)&&\j=\sqrt{\k}\t\nonu
\del\bar{\vf}=\frac{2}{\sqrt{\k+1}}(\hat{E}^0-
\hat{U}^0) && \bar{\j}=\sqrt{\k}\bar{\t}.
\eea
This provides us with a realization where $G_+$ and $G_-$ are treated on the
same footing. This is
the so-called symmetric realization of the $N=2$ algebra: both $G_+$ and $G_-$
are given by
composite operators. In order to obtain a stringy interpretation of the
reduction, we want to identify
$G_+$ with the BRST current and $G_-$ with the Virasoro anti-ghost, so a single
field instead of a
composite.
To achieve this we consider a different grading, namely according to the
eigenvalues of
$ \frac 1 2 \mbox{ad}_{e_0}+ \mbox{ad}_{u_0}$. We obtain for the gradings of
the
various
currents:\\[.1cm] \begin{center}
{ \begin{tabular}{|c||c|c|c||c||c|c||c|c||}\hline $
$&$E^\pp$&$E^0$&$E^=$&$U^0$&$F^{+}$
&$F^{-}$&$\bar{F}^{+}$&$ \bar{F}^{-}$ \\ \hline
${\it grade}$&$1$&$0$&$-1$&$0$&$3/2$&$1/2$&$-1/2$&$-3/2$\\ \hline
\end{tabular}}\ .\\[1cm]
\end{center}
We follow quite the same procedure as before, but whenever we refer to the
grading on the Lie algebra,
we always imply it to be the grading induced by $\frac 1 2 \mbox{ad}_{e_0}+
\mbox{ad}_{u_0}$.
Again we constrain the strictly negatively graded part of the algebra:
\bea
\Pi_{<0} J =\frac \k 2\left(e_= + \psi \bar{g}_+ \right). \eea
The appearance of the auxiliary field $\psi$ is understood as follows. The
current $\bar{F}^+$ is a
highest $sl(2)$ weight and will become the leading term of a conformal
current \cite{ST}.
But on the same token, $\bar{F}^+$ has a negative grading, so it has to be
constrained. Thus we need
to constrain it in a non-singular way, {\it i.e.} by putting it equal to an
auxiliary field which is inert under
the gauge transformations.
The action which reproduces the constraints is easily obtained: \bea
{\cal S}=\k S^-[g]+\frac{1}{\pi x}\int str\, A(J-\frac \k 2 e_=-\frac \k 2 \Psi
)
+\frac{\k}{2\p x}\int str \Psi \bdel\bar{\Psi}, \eea
where
\bea
A&=&A^\pp e_\pp+A^+g_++A^-g_-,\nonu
\Psi&=&\psi\bar{g}_+,\qquad
\bar{\Psi}=\bar{\psi}g_-.
\eea
The gauge invariance is parametrized by $h=\exp \h$, $\h\in\P_{>0} sl(2|1)$ or
$\h=\h^\pp e_\pp
+\h^+g_++\h^-g_-$, \bea
g&\rightarrow& g'=hg,\nonu
A&\rightarrow& A'=\bdel h h^{-1} + h A h^{-1},\nonu \bar{\Psi}&\rightarrow&
\bar{\Psi}'=\bar{\Psi} + \h^-g_-. \eea
One immediately sees that the combined requirements of gauge
invariance and the existence of
a non-degenerate highest weight gauge, requires the introduction of the
field $\bar{\psi}$ conjugate
to $\psi$, even if it was not needed for the constraints.
As before, the gauge choice is $A=0$. Introducing ghosts $c = c^\pp
e_\pp+\g^+g_++\g^-g_-
\in \P_{>0} sl(2|1)$ and anti-ghosts $b = b^= e_= + \bar{\b}^+ \bar{g}_+
+\bar{\b}^- \bar{g}_-
\in \P_{<0} sl(2|1)$, we get the gauge fixed action: \bea
{\cal S}_{gf}=\k S^- [g] +\frac{\k}{2\p x}\int str \Psi \bdel\bar{\Psi}
+\frac{1}{2\p x}\int str b\bdel c, \eea
and the BRST charge
\bea
Q_{HR}=\frac{1}{4\p i x}\oint str \left\{ c\left( J-\frac \k 2 e_= -\frac \k 2
\Psi + \frac 1 2 J_{gh} \right)
\right\}. \eea
Again, the affine symmetry of the WZW model breaks down to an $N=2$
superconformal symmetry
whose generators the generators of the cohomology ${\cal H}^*({\cal
A},Q_{HR})$,
where $\ca$ is
the algebra generated by $\{ b, \hat{J}=J+ J_{gh}, \psi, \bar{\psi}, c\}$ and
all normal ordered products
of these fields and their derivatives. We are now in the position to use
exactly
the same methods, {\it i.e.}
spectral sequence techniques, to solve for the cohomology provided we consider
the double grading
$(m,n)$ where $m$ is induced by the action of $\frac 1 2 \mbox{ad}_{e_0}+
\mbox{ad}_{u_0}$ and
$m+n$ is the ghost number. We assign grading $(0,0)$ to the auxiliary fields
$\psi$ and $\bar{\psi}$.
Applying the methods developed in \cite{ST}, one arrives at
\bea
T_{N=2}&=& \frac{2\k}{\k+1}\left(\hat{E}^\pp+ \psi \hat{{F}}^- -\frac 2 \k
\hat{U}^0 \hat{U}^0
+\frac 2 \k \hat{E}^0 \hat{E}^0\right.\nonu
&&\left. -\del \hat{E}^0-\frac{1}{\k}\del \hat{U}^0 -\frac{\k +1 }{4}\left(
3\psi\del\bar{\psi}
+\del\psi \bar{\psi}\right)\right)\nonu
G_+&=& \frac{2\k^2}{1+\k}\left( \hat{F}^+ + \hat{E}^\pp\bar{\psi}- \frac 2 \k (
\hat{E}^0
+ \hat{U}^0)\hat{F}^- + \hat{F}^-\bar{\psi} \psi + \del \hat{F}^- \right.\nonu
&&\left.
-\frac{(\k+1)(2\k+1)}{4\k}\del^2 \bar{\psi} + \frac 2 \k ( \hat{E}^0\hat{E}^0 -
\hat{U}^0\hat{U}^0 )
\bar{\psi}\right.\nonu &&\left. - \del (\hat{E}^0-\hat{U}^0) \bar{\psi}
-\frac{1+\k}{2}\psi \del \bar{\psi}
\bar{\psi} + \frac{2(1+\k)}{\k} \del \bar{\psi}\hat{U}^0\right) \nonu
G_-&=&\psi\nonu
U&=&-4\left( \hat{U}^0+\frac \k 4 \psi\bar{\psi}\right) , \eea
which satisfies the $N=2$ superconformal algebra with $c_{N=2}= -3(1+2\k)$.
Again, the Miura transform is given by the algebra isomorphism $X\rightarrow
\P_{(0,0)}X$, where
$X$ stands for the conformal currents. Together with the OPE's
$\hat{E}_0(z_1)\hat{E}_0(z_2)=- \hat{U}_0(z_1)\hat{U}_0(z_2) = (\k + 1)/8 \,
z_{12}^{-2}$
and $\psi(z_1) \bar{\psi}(z_2) = 1/\k \, z^{-1}_{12}$ gives the desired
realization
of the $N=2$ algebra.
Indeed, identifying $B \equiv \psi$, $C \equiv \k \bar{\psi}$, $\del \varphi_L
\equiv \sqrt{8 / (\k +1)} \hat{U}_0$
and $\del \varphi_m \equiv i\sqrt{8 / (\k +1)} \hat{E}_0$, and
\bea
T_m=-\frac 1 2 \del\vf_m\del\vf_m + i \frac{\k}{\sqrt{2(\k+1)}}\del^2\vf_m,
\eea
precisely reproduces, upon twisting, the non-critical string theory discussed
at the beginning of this section
with
\bea
c_m=1-6\left (\sqrt{\k+1}-\frac{1}{\sqrt{\k+1}}
\right)^2.
\eea
It is interesting to note that if we take for the matter sector a reduction of
the $sl(2)$ WZW model, one
gets for $c$ in terms of the level $\k_{sl(2)}$ of the underlying $sl(2)$
WZW model precisely the previous expression for $c_m$, provided one identifies
$\k + 1$
with $\k_{sl(2)}+2$. The shift can be udenrstood as an additional
ghostcontribution to the $sl(2)$ central charge.
This is not a stand alone case. This program has been carried out in a few
other
cases as well: strings
with $N$ supersymmetries have the Knizhnik-Bershadsky $SO(N+2)$ superconformal
symmetry and
are obtained from the reduction of $OSp(N+2|2)$ \cite{BLLS}. Classically, it
was
shown in \cite{BLNW}
that $W_n$ strings have an $N=2$ $W_n$ algebra and they are obtained from a
reduction of $Sl(n|n-1)$.
We will now turn to a detailed investigation of this construction. In the next
sections we analyze the embeddings
of $sl(2|1)$  in supergroups after which we come back to the construction of
stringtheories.
\sect{Classification of \ls\ embeddings into Lie superalgebras \label{class}}

\indent

Let \cg\ be a Lie superalgebra. We want to classify its \ls\ \ssa s. As \ls\
admits an \osp\ (principal) \ssa,
the classification of \ls\ embeddings is a subclass of \osp\ ones. Let us first
recall the two
fundamental theorems concerning the classification of \osp\ \ssa
\cite{DRS,FRS}:

\subsection{$osp(1|2)$ embeddings: a reminder}

\indent

{\it {Principal \osp\  embeddings in superalgebras}

The only simple superalgebras that admit a principal \osp-embedding are the
$s\ell(n|n\pm1)$, $osp(2n|2n\pm1)$, $osp(2n|2n)$, $osp(2n|2n+2)$, or
$D(2,1;\alpha)$
superalgebras. They all admit a totally fermionic simple roots
system\footnote{This necessary
condition is almost sufficient, since only the $s\ell(n|n)$ superalgebra has a
totally fermionic simple
roots system while not admitting a principal \osp.}
}

We recall that the principal \osp-subalgebra of a superalgebra \cg\ possesses
as simple fermionic root
generator the sum of all the \cg\ simple fermionic root generators. It is
maximal in \cg\ (i.e. the only
superalgebra that contains the principal \osp\ in \cg\ is \cg\ itself). On the
opposite, a superalgebra
\ch\ is regular in \cg\ when its root generators are root generators for \cg:
the regular \osp\ in \cg\ is
the "smallest" subalgebra of \cg.

\indent

{\it {Classification of \osp\ embeddings}

Any \osp-embedding in a simple Lie suparalgebra \cg\ can be considered as the
principal \osp\
of a regular \cg-sub-superalgebra (of the type given just above), up to the
following exceptions:
\begin{itemize}
\item For \cg= $osp(2n\pm2|2n)$ with $n\geq2$, besides the principal embeddings
described above,
there exists also \osp\ sub-superalgebras associated to the singular embeddings
$osp(2k\pm1|2k)\oplus osp(2n-2k\pm1|2n-2k)$ with $1\leq k\leq
\left[\frac{n-1}{2}\right]$.
\item For \cg= $osp(2n|2n)$ with $n\geq2$, besides the principal embeddings
described above,
there exists also \osp\ sub-superalgebras associated to the singular embeddings
$osp(2k\pm1|2k)\oplus osp(2n-2k\mp1|2n-2k)$ with $1\leq k\leq
\left[\frac{n-2}{2}\right]$.
\end{itemize}
}

\subsection{$s\ell(1|2)$ embeddings}

\indent

As already mentionned, any \ls\ \ssa\ provide an \osp-embedding. Hence, the
classification of
\ls-embeddings
will also be associated to some of the above \cg-\ssa(s). Let us be precise, we
have:
\begin{theor}\label{theo1}
Let \cg\ be a superalgebra. Any \ls\ embedding into \cg\ can be seen as the
principal \ssa\ of a (sum
of) regular
$s\ell(p|p\pm1)$ \ssa(s) of \cg, except in the case of $osp(m|2n)$ ($m>1$),
$F(4)$ and $D(2,1;\alpha)$
where the (sum of) regular $osp(2|2)$ has also to be considered\footnote{We
make a distinction
between the $osp(2|2)$ and \ls\ superalgebras: these two superalgebras are
isomorphic, but not the
associated supergroups. The smallest non-trivial representation for $osp(2|2)$
is 4-dimensionnal,
while it is 3-dimensionnal for \ls.}.
\end{theor}

This theorem will be proved using some lemmas that we introduce now.

\begin{lem}\label{lem0}
The principal \osp\ \ssa\ of $s\ell(n|n\pm1)$ can be "enlarged" to a principal
\ls\ \ssa.
\end{lem}

\prf It is obvious from the construction of the principal \osp \ssa, as it has
being presented in \cite{DRS}.

Using this lemma, one can immediately obtain a classification of \ls\ \ssa\
into $s\ell(m|n)$
superalgebras:

\begin{lem}\label{lem1}
For $\cg=s\ell(m|n)$, the \osp\ embeddings classify the \ls\ ones. \end{lem}

\prf One already knows that in $s\ell(m|n)$, the \osp\ embeddings are
associated to (sums of) $s\ell(p|p\pm 1)$ \ssa s. But from lemma \ref{lem0}, we
know that these superalgebras admit a principal \ls. As two different \osp\
\ssa s cannot be principal in the same \ls\ \ssa, we deduce that each \osp\ is
associated to a different \ls.

\begin{lem}\label{lem2}
Let \cg=$osp(2n\pm2|2n)$ or $osp(2n|2n)$. We consider an \osp-embedding
classified by a
\underline{singular} \ssa\ in \cg\ (see classification of \osp\ embeddings).
Then, there is no \ls\ in
\cg\ that contains the \osp\ under consideration. \end{lem}

\prf The proof relies on the decomposition of the adjoint of \cg\ into
\osp\ representations $R_{j}$. It has been given in the reference \cite{FRS}
and in each
case\footnote{Be careful of a misprint about the boundary values for $k$ in the
reference \cite{FRS}} it is easy to see that there is no $R_{1/2}$
representation in the \cg-adjoint.
As an \ls\ decomposes into $R_1\oplus R_{1/2}$ under its \osp\ \ssa\ (see
below), the lemma
\ref{lem2} is clear.

\begin{lem}\label{prop1}
Let \ch\ be a \underline{regular} \ssa\ of \cg\ defining an \osp\ embedding in
\cg.
Let $\{\alpha_i\}$ be the set of fermionic simple roots of \ch, and
$\{\beta_j\}$ the
set of simple roots of $\ch_0$, the bosonic part of \ch. We suppose that the
principal
\osp\ \ssa\ of \ch\ can be enlarged in \cg\ to a \ls\ \ssa, and we denote by
$B$ the \ls-Cartan
generator that commutes with $\su\subset\ls$.

Then, $[B, E_{\pm\beta_j}]=0$ and
$[B,F_{\pm\alpha_i}]= \pm b_i\ F_{\pm\alpha_i}$ with $b_i\neq0$.
\end{lem}

\prf \ch\ is regular in \cg, thus the root-generators of \ch\ are
root-generators
(i.e. eigenvector under any Cartan generator) of \cg. As $B$ is a Cartan
generator, we deduce that
\be
{[B, F_{\pm\alpha_i}]= \pm b_i F_{\pm\alpha_i}} \label{eq.BF}
\ee
The same argument for the bosonic part
$\ch_0$ leads to
\be
{[B, E_{\pm\beta_j}]=\pm y_j E_{\pm\beta_j}} \label{eq.BE}
\ee
Now, denoting by $E_\pm$ and $F_\pm$ the root-generators of \osp\ and by
$F_{\pm}^{\pm}$ the fermionic roots of \ls, we have
\be\begin{array}{l}
F_\pm=F_\pm^++F_\pm^- \\
{[B,E_\pm]}=0 \mb{and} [B,F_\pm]=F_\pm^+-F_\pm^-, \label{toto}
\end{array}
\ee
 But we
can choose $E_+=\sum_jE_{\beta_j}$ and $F_+=\sum_iF_{\alpha_i}$, where
the sums run over all the simple roots of $\ch_0$ and \ch\ respect. Applying
the formulae
(\ref{eq.BF}-\ref{eq.BE}) to (\ref{toto}) leads to $y_j=0$ and $b_i\neq0$.

\begin{lem}\label{lem3}
Let \ch\ be a \cg-regular \ssa\ which possesses a principal \osp. If the \osp\
can be enlarged (in \cg) to a
\ls\ \ssa, then the totally fermionic simple root basis of \ch\ does not
contain an \osp-type root ("black
root") \end{lem}

\prf We proove this lemma {\it ad absurdum}. Let $\alpha_0$ be the "black"
simple root of \ch. Then, $2\alpha_0$ is also a (bosonic) root (in \ch). From
lemma \ref{prop1},
we have $[B,E_{2\alpha_0}]=0$ and $[B,F_{\alpha_0}]=b_0 F_{\alpha_0}$ with
$b_0\neq0$.
These two commutation relations are clearly incompatible, as it can be seen
from the Jacobi identity
$(B,F_{\alpha_0},F_{\alpha_0})$. Thus the black root $\alpha_0$ does not exist
if the \ls-generator $B$ does.

\begin{lem}\label{lem4}
Under the same conditions as in lemma \ref{lem3}, the totally fermionic Dynkin
diagramm of \ch\ does
not contain any "triangle" of roots:

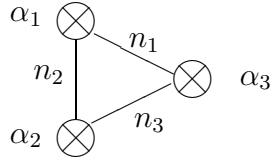
\begin{center}
\begin{picture}(150,60)
\put(60,8){\circle{14}}
\put(35,6){$\alpha_2$}
\put(55,3){\line( 1, 1){10}}
\put(55,13){\line( 1, -1){10}}
\put(60,52.1){\circle{14}}
\put(35,52.1){$\alpha_1$}
\put(55,47){\line( 1, 1){10}}
\put(55,57){\line( 1, -1){10}}
\put(44,30){$n_2$}
%\put(63,15){\line( 0, 1){ 30}}
\put(60,15){\line( 0, 1){ 30}}
%\put(57,15){\line( 0, 1){ 30}}
%
\put(82,12){$n_3$}
\put(66,13.5){\line( 2, 1){ 30}}
\put(80,42){$n_1$}
\put(67,47){\line( 2, -1){ 30}}
%\put(68,49){\line( 2, -1){ 30}}
%
\put(104,30){\circle{14}}
\put(122,28){$\alpha_3$}
 \put(99,25){\line( 1, 1){10}}
\put(99,35){\line( 1, -1){10}}
 \end{picture}
 \end{center}

where $n_i\neq0$ ($i=1,2,3)$ represent the non-vanishing numbers of lines.
 \end{lem}

\prf The existence of the triangle implies that $\alpha_1+\alpha_2$,
$\alpha_2+\alpha_3$ and $\alpha_3+\alpha_1$ are all bosonic simple roots (for
$\ch_0$). Then, the
lemma \ref{prop1}, with the help of the Jacobi
identities $(B,F_{\alpha_i},F_{\alpha_i})$, $i=1,2,3$, constrains the
$B$-eigenvalues $b_1$, $b_2$ and $b_3$ of $F_{\alpha_1}$, $F_{\alpha_2}$
and $F_{\alpha_3}$ to satisfy $b_1+b_2=0$, $b_2+b_3=0$ and $b_3+b_1=0$.
Again, it is incompatible with $b_i\neq0$.

Note that the proof relies only on the fact that the sum of any couple of roots
in the triangle is a root:
thus, this lemma also excludes $D(2,1;\alpha)$ as a candidate for \ls\
classification\footnote{We
recall that although there exists a $N=4$ superconformal algebra based on
$D(2,1;\alpha)$, it
is obtained from the Hamiltonian reduction of $D(2,1;\alpha)$ w.r.t. a
\underline{regular}
$osp(2|2)$ subalgebra, and not a (possible) principal \ls-embedding.}.

\indent

Gathering the lemmas \ref{lem3} and \ref{lem4}, we see that for any
superalgebra \cg, the
\underline{regular} \ssa s classifying \ls\
embeddings possess a totally
fermionic Dynkin diagramm without any black root, nor triangle. Looking at the
list given at the
beginning of this section, it is clear
that only the (sums of) superalgebras $s\ell(n|n\pm1)$ and $osp(2|2)$ obey to
these
rules. The $osp(2|2)$ \ssa\ is isomorphic to $sl(1|2)$ but gives a different
decomposition of the
fundamental representation (see below). The only cases where
$osp(2|2)$ appears are\footnote{Note that, for $G(3)$,
the classification of $osp(1|2)$ embeddings done in
\cite{FRS}, table 15, mention the regular $osp(1|2)=B(0,1)$ and not the
$osp(1|2)$ principal in
$osp(2|2)$ because  these two embeddings are equivalent.}
$\cg=osp(m|2n)$ (with $m>1$), $G(3)$, $F(4)$ and $D(2,1;\alpha)$.
The lemma \ref{lem0} ensures that the other \ssa s can indeed be associated to
a \ls\ \ssa.
Moreover, the only case (including exceptional superalgebras) where singular
embeddings are
required (in the classification of \osp\ \ssa s) have been treated in the lemma
\ref{lem2}.
So the theorem \ref{theo1} is proved for all the simple Lie superalgebras. It
extends trivially to
sums of simple Lie superalgebras.

\sect{Decomposition of Lie superalgebras w.r.t. \ls}
\subsection{Summary on \ls\ representations \label{sl12rep}}

\indent

Denoting by $E_\pm$ and $H$ the \su\ generators and $B$ the $g\ell(1)$
generator that commutes with \su\ in \ls, the states of an \ls-irreducible
representation will be
classified according to their $H$-"isospin" $j$ and "baryon number" $b$. One
can distinguish:

\indent

$\bullet$ {\bf Typical representations $(b,j)$ with $b\neq\pm j$}

\indent

For $j\geq1$, they are made with the $(2j+1)$ states of the \su-representation
$\cd_j$ with a $B$-eigenvalue $b$, together with the states of two
$\cd_{j-1/2}$ representations
of $B$-eigenvalue $b\pm\half$ respectively, and finally the states of the
$\cd_{j-1}$-representation
with $B$-eigenvalue $b$. Reducing the representation $(b,j)$ w.r.t. its
$\su\times g\ell(1)$ subalgebra,
we will note:
\be
(b,j)= |b,j>\oplus |b-\half,j-\half>\oplus |b+\half,j-\half>\oplus |b,j-1>
\ee

For $j=\half$, the representation $(b,\half)$ reads
\be
(b,\half)= |b,\half>\oplus |b-\half,0>\oplus |b+\half,0>
\ee

The dimension of a typical representation $(b,j)$ is $8j$.

\indent

$\bullet$ {\bf Atypical representations $(b=j,j)$ and $(b=-j,j)$ with $j\geq0$}

\indent

For $j\neq0$, and using the same notations as above, we have
\be
(\pm j,j)= |\pm j,j>\oplus |\pm (j+\half), j-\half>
\ee

The $j=0$ atypical representation is just the trivial representation.

The dimension of the atypical representations $(\pm j,j)$ is $4j+1$.

\indent

We want to emphasize that the sign of the $U(1)$ charge in an atypical
representation has no real meaning: the two representations $(\pm j,j)$
are related by an outer automophism of the $sl(1|2)$ algebra. We will come back
later on this
point that has some consequence in the decomposition of the adjoint
representation w.r.t. \ls.

\indent

Note that if one decomposes the \ls-representations with respect to the
\osp-\ssa\ of \ls, the typical
representation $(b,j)$ corresponds to the sum of two \osp-representations
$R_j\oplus R_{j-1/2}$,
while the atypical representation $(\pm j,j)$ is just a $R_j$ representation.

Let us also remark that the two \ls-Casimir operators are zero in an atypical
representation.

\indent

Finally, we want to stress that the product of two \ls-irreducible
representations is \underline{not}
always completely reducible.
%Hopefully, we do not have this pathology in the products we consider
% after.

\subsection{Products of \ls-representations \label{prodRep}}

\indent

Our aim is to decompose the adjoint representation of a simple Lie superalgebra
\cg\ into representations
of \ls\ considered as a \ssa\ of \cg. Following the techniques used for the
decomposition of \cg\ into
\osp-representations, we will start by decomposing the \cg-fundamental
representation. Then, performing
the product of this fundamental representation by its contragredient, we will
obtained the
desired decomposition. One will be helped by the following formulae \cite{SNR}:
\bea
(\pm j,j)\times(\pm k,k) &=& (\pm (j+k), j+k)
\oplus_{l=|j-k|+\half}^{j+k-\half}
(\pm (j+k+\half)\ ,\ l) \\
&=& (\pm (j+k), j+k)\oplus (\pm (j+k+\half), j+k-\half)\oplus \nonumber\\
&& \oplus
(\pm (j+k+\half), j+k-\frac{3}{2})\oplus \dots \oplus (\pm (j+k+\half),
|j-k|+\frac{3}{2}) \oplus \nonumber\\ && \oplus
(\pm (j+k+\half), |j-k|+\half)
\ \ \ \ \ \ \ \ \forall\ \ \ j,k\geq0 \nonumber \ena
If we consider only products of atypical representations
we already know that they are decomposable into sum of irreducible
\ls-representations (since they
are of the type S$_{\pm}$ introduced in \cite{SNR}). Thus, we can focus on the
$s\ell(2)\oplus g\ell(1)$
part to deduce informations about \ls-representations. Then, using the
$\su\times g\ell(1)$ decomposition
given above, it is also possible
to compute:
\be
(j,j)\times(-k,k) = (j-k,j+k)\oplus
(j-k, j+k-1)\oplus \dots\oplus (j-k, |j-k|) \ee

As examples, we have
\be
\begin{array}{l}
(j,j)\times(-j,j) =(0,2j)\oplus (0,2j-1)\oplus (0,2j-2)\oplus\dots \oplus (0,0)
\\ \\
(\pm\half,\half)\times (\pm\half,\half)
=(\pm1,1)\oplus (\pm\frac{3}{2},\half) \mb{while} (\half,\half)\times
(-\half,\half) =
(0,1)\oplus (0,0)
\end{array}
\ee

Considering indecomposable products, we have for instance \cite{SNR}
$(0,\half)\times(0,\half)$: we will come back extensively on this point in
section \ref{osp22}.

\indent

We will also need to select the (anti)symmetric part of the product of
representations. For brievety, we will note $[(b,j)]^2_S$ and $[(b,j)]^2_A$ the
symmetric
and antisymmetric part of the product $(b,j)\times (b,j)$. Using the rules
given in \cite{FRS} for
$s\ell(2)\oplus g\ell(1)$ representations, it is easy to deduce:

\be
\begin{array}{l}
\mb{For}\ m \in \N  \\
\Asym{(m,m)\oplus (-m,m)}^2\ =\ \oplus_{j=0}^{2m} (0,j)\ \oplus_{j=1}^{m}\
(2m+\half, 2j-\half)\
\oplus_{j=1}^{m}\ (-(2m+\half), 2j-\half) \\
\\
\Asym{(m+\half,m+\half) \oplus (-(m+\half), m+\half)}^2 =
\begin{array}{l}
\oplus_{j=0}^{2m+1} (0,j)\ \oplus_{j=0}^{m}\ (2m+\frac{3}{2}, 2j+\half)\
\oplus\\
 \oplus_{j=0}^{m}\ (-(2m+\frac{3}{2}), 2j+\half)
\end{array} \\
\\
\Sym{(m,m)\oplus (-m,m)}^2\ =\
\begin{array}{l}
\oplus_{j=0}^{2m}(0,j)\ \oplus_{j=0}^{m-1}\ (2m+\half, 2j+\half)
\oplus_{j=0}^{m-1}\
(-(2m+\half), 2j+\half) \\
 \oplus (2m,2m) \oplus(-2m,2m)
\end{array}  \\
\\
\Sym{(m+\half,m+\half)\oplus (-(m+\half),m+\half)}^2\ =\
\begin{array}{l}
 \oplus_{j=0}^{2m+1} (0,j)\ \oplus_{j=1}^{m}\ (2m+\frac{3}{2}, 2j-\half)\
\oplus\\
 \oplus_{j=1}^{m}\ (-(2m+\frac{3}{2}), 2j-\half) \oplus\\
\oplus (2m+1,2m+1)\oplus (-(2m+1), 2m+1) \end{array}
\end{array}
\ee
together with
\beano
\mbox{For}\ \ \ j \in \N/2 && \\
\Asym{n(\pm j,j)\times n(\pm j,j)} &=& \frac{n(n+1)}{2} \Asym{(\pm j,j)\times
(\pm j,j)}
\oplus\frac{n(n-1)}{2} \Sym{(\pm j,j)\times (\pm j,j)} \\
\Sym{n(\pm j,j)\times n(\pm j,j)} &=& \frac{n(n+1)}{2} \Sym{(\pm j,j)\times
(\pm j,j)}
\oplus\frac{n(n-1)}{2} \Asym{(\pm j,j)\times (\pm j,j)}\\
\enano
We will also use
\be
\begin{array}{l}
\mb{For}\ j_1,j_2 \in \Z/2 \\
\Asym{ (j_1,|j_1|)\oplus (j_2,|j_2|) }^2 \ =\
\begin{array}{l}
\Asym{(j_1,|j_1|) \times (j_1,|j_1|)} \oplus\Asym{(j_2,|j_2|) \times
(j_2,|j_2|)} \oplus \\
 \oplus (j_1,|j_1|)\times (j_2,|j_2|)
\end{array}
 \\ \\
\Sym{ (j_1,|j_1|)\oplus (j_2,|j_2|) }^2\ {=}\
\begin{array}{l}
 \Sym{(j_1,|j_1|) \times (j_1,|j_1|)} \oplus\Sym{(j_2,|j_2|) \times
(j_2,|j_2|)}\oplus \\
 \oplus(j_1,|j_1|)\times (j_2,|j_2|) \end{array}
\end{array}
\ee

\subsection{Superalgebras fundamental representations \label{fundRep}}

\indent

The techniques for the decomposition of the fundamental representations is
the same as in \cite{FRS}. We start with a superalgebra \cg\ that we want
to reduce w.r.t. a given \ls-subalgebra, defined through its principal
embedding into a \cg-subalgebra \ch. Decomposing \ch\ into its simple parts
$\ch_i$:
$\ch=\oplus_{i}\ \ch_i$, we associate to each type of $\ch_i$ and each type of
superalgebra \cg,
a \ls-representation. Then, we
sum these different \ls-representations, and eventually complete this sum
by trivial representations, in such a way that the dimension of the
\cg-fundamental is recovered.

For $\cg=s\ell(m|n)$ superalgebras, we will get a $(\pm
\frac{p}{2},\frac{p}{2})$ atypical
representation for each $s\ell(p+1|p)$ \ssa\ and a $(\pm
\frac{p}{2},\frac{p}{2})^\pi$
representation for each $s\ell(p|p+1)$ \ssa. We use
the superscript $\pi$ to distinguish the
$(\pm \frac{p}{2},\frac{p}{2})$ representations coming from these two types of
superalgebras.
When the trivial representation occurs, it will be denoted
$(0,0)$ if it is in the $s\ell(m)$ fundamental representation, and $(0,0)^\pi$
if it is in the $s\ell(n)$
one.
This superscript $\pi$ will have deep consequences on
the spin structure of the resulting adjoint decomposition. We repeat that the
sign of the $U(1)$ charge
is meaningless (see section \ref{adjRep})

For $\cg=osp(m|2n)$ superalgebras, we get a sum $(\frac{p}{2},\frac{p}{2})
\oplus (-\frac{p}{2},\frac{p}{2})$
of atypical representations (here the sign of the $U(1)$ charge has been fixed
by the reality condition)
for each $s\ell(p+1|p)$ \ssa\ and a sum
$(\frac{p}{2},\frac{p}{2})^\pi\oplus (-\frac{p}{2},\frac{p}{2})^\pi$
of representations for each $s\ell(p|p+1)$ \ssa. Again, the trivial
representation will be denoted
$(0,0)$ if it is in the $O(m)$ fundamental representation, and $(0,0)^\pi$ if
it stands in the $sp(2n)$
one.

\subsection{\ls-decomposition of the adjoint representation\label{adjRep}}

\indent

The rules given in the section \ref{prodRep} do not indicate the statistics
of the representation. In other words, when we obtain a representation
$(b,j)$ in the adjoint of \cg, we have not yet specified whether the $\cd_{j}$,
$\cd_{j-1/2}$, and
$\cd_{j-1}$ representation are associated to commuting or anti-commuting
generators of \cg. A natural
statistics associate (anti-)commuting generators to (half-)integers $j$, but
there
are some cases where it is the opposite. To distinguish these two (very
different) cases, we will note
{\it in the adjoint representation}
with a prime $(b,j)'$
the representation with "unsual" statistics, keeping the form $(b,j)$ for the
representations with usual
statistics.

Then, the rules to distinguish the two kinds of representations are the
same as the ones given for $osp(1|2)$ representations, i.e. \bea
(b_1,j_1)\times (b_2,j_2) &=& \left\{
\begin{array}{l}
\oplus_{b_3,j_3} (b_3,j_3) \mb{if} j_1+j_2\in\Z \\ \oplus_{b_3,j_3} (b_3,j_3)'
\mb{if} j_1+j_2\in\half+\Z \end{array}
\right. \\
(b_1,j_1)^\pi\times (b_2,j_2)^\pi &=& \left\{ \begin{array}{l}
\oplus_{b_3,j_3} (b_3,j_3) \mb{if} j_1+j_2\in\Z \\ \oplus_{b_3,j_3} (b_3,j_3)'
\mb{if} j_1+j_2\in\half+\Z \end{array}
\right. \\
(b_1,j_1)\times (b_2,j_2)^\pi &=& \left\{ \begin{array}{l}
\oplus_{b_3,j_3} (b_3,j_3)' \mb{if} j_1+j_2\in\Z \\ \oplus_{b_3,j_3} (b_3,j_3)
\mb{if} j_1+j_2\in\half+\Z \end{array}
\right.
\ena

Using the rules given above, it is now easy to get the decomposition of the
adjoint from the product
$F\times \bar{F}$, where $F$ is the fundamental representation (deomposed into
$s\ell(1|2)$
representations):

For $s\ell(m|n)$ superalgebras, the decomposition of the adjoint representation
will be
$(F\times \bar{F})- (0,0)$, the rules for the product being given in section
\ref{prodRep},
with the property $\overline{\r{p}}=(-p,p)$.

Note that the choice between $(p_1,p_1)\oplus(p_2,p_2)$ or
$(p_1,p_1)\oplus(-p_2,p_2)$ leads to
very different adjoint decomposition. However, these decompositions are
equivalent. Indeed,
the change $(p_2,p_2)\ \rightarrow\ (-p_2,p_2)$ correspond to the following
changes (algebra isomorphism)
in the \ls-generators: $Y\rightarrow -Y$; $F_{+-}\rightarrow F_{++}$;
$F_{++}\rightarrow F_{+-}$;
$F_{--}\rightarrow -F_{-+}$; $F_{-+}\rightarrow -F_{--}$; $E_\pm\rightarrow
E_\pm$;
$H\rightarrow H$. This is clearly just a choice of normalisation and thus
correspond to equivalent
\ls-subalgebras. Note that in the series $s\ell(2)$, \osp, \ls, the \ls\
superalgebra is the first (super)algebra possessing an outer automorphism: this
explains why
these "multiple adjoint decompositions" have not being encountered when
studying \osp\ and
$s\ell(2)$ decompositions of
(super)algebras. In the next section, we show explicitely on an example
how things are going on. It should be clear to the reader that "multiple
\ls-decomposition" will occur
only where more than one non-trivial atypical representation are involved in
the fundamental representation.
\indent

For $osp(m|2n)$ algebras, the adjoint decomposition will be given by the
antisymmetric
product of $(p,p)$ representation, plus the symmetric product of $\rpi{p}$,
plus {\bf once} the
products of the $\r{p}$'s by the $\rpi{q}$'s
representations.

\subsection{Case of $osp(2|2)$ subalgebras and indecomposable products
\label{osp22}}

\indent

Up to now, we have studied the embeddings of $s\ell(1|2)$ into superalgebras.
However, $s\ell(1|2)$ is
isomorphic to another superalgebra, namely $osp(2|2)$. As they are isomorphic,
one could think that one
has not to distinguish them. However, already at the level of representations,
it is clear that these two
superalgebras are distinct, since, for instance, $osp(2|2)$ has only real
representations, while $s\ell(1|2)$
has complex ones. This distinction appears also here when decomposing the
fundamental of an $osp(m|2n)$
superalgebra w.r.t. $osp(2|2)$.
If the sub-superalgebra is an $osp(2|2)$ superalgebra (instead of a
$s\ell(1|2)$ one) we will get one
$\ropi{\half}$ representation in the
fundamental. Note that the distinction between the two isomorphic superalgebras
$osp(2|2)$ and \ls\ is of
the same type as the one introduced in \cite{FRS} to
distinguish between the \su-decomposition coming from the algebras $A_1$ and
$C_1$ (in symplectic algebras), $D_2$ and $2A_1$, or $D_3$ and $A_3$ (in
orthogonal algebras).

Thus, considering the decomposition of $osp(m|2n)$ superalgebras, we have to
add the cases where one
or several $osp(2|2)$ appear. For each $osp(2|2)$ subalgebra, we will have a
$(0,\half)^\pi$ representation
in the fundamental of $osp(m|2n)$. Then, the decomposition of the adjoint
representation will be obtained
with the same rules as given in previous section.

However, note that the product of two $(0,\half)$ representations is not
completely reducible.
More precisely, the symmetric part of $(0,\half)\times(0,\half)$ contains a
$(0,1)$ representation, while
the antisymmetric part is non fully reducible: from the $s\ell(2)\oplus
g\ell(1)$ decomposition this part
{\it looks} like $(\half,\half)\oplus (-\half,\half)\oplus 2(0,0)$, but one
verifies that one of the two
$D_0(0)$ generators is obtained from both the "$(\half,\half)$" and
"$(-\half,\half)$" parts by application
of negative root generators.
Thus, apart from a $(0,0)$ representation, the antisymmetric part of this
product is non fully reducible.
Below, we will keep the notation $[(0,\half)]^2_A$ for the indecomposable part
(plus a trivial
representation).
\be
\Sym{\ro{\half}}^2= \ro{1} \mb{while}
\ro{\half}\times\ro{\half}=\ro{1}\oplus\rinde
\ee
Therefore, when one decomposes $osp(m|2n)$ superalgebras w.r.t. the diagonal of
several regular
$osp(2|2)$ subalgebras, one will get an {\it non fully reducible part}.
Although this fact seems quite
intringuing, one has to remind that most of the representations of Lie
superalgebras are
{\it non fully reducible} \cite{SNR}.
Fortunately, the product of $\ropi{\half}$ with an atypical representation is
 reducible:
\be
(\pm j,j)\times(0,\half) =(\pm j,j+\half) \oplus (\pm j,j-\half)  \oplus
(\pm j+\half,j)\oplus (\pm j-\half,j)
\ee

As an example, let us consider the reduction of $osp(4|4)$ w.r.t. $2\
osp(2|2)$. The fundamental
reads \be
\underline{4}= 2\ropi{\half}
\ee
while the adjoint is
\be
\left[ 2\ro{\half} \right]^2_{\bf S}= \frac{2.3}{2} [(0,\half)]^2_{\bf S}
\oplus \frac{2.1}{2}
[(0,\half)]^2_{\bf A}
=3\ (0,1) \oplus  [(0,\half)]^2_{\bf A} \label{dodo}
\ee
We have explicitely checked that the decomposition is really the one given in
(\ref{dodo}).

\subsection{Example \label{ex}}

Let us treat one of the simplest examples where the problem of multiple
$s\ell(2|1)$-decomposition occurs, namely the reduction $s\ell(4|3)$ w.r.t.
the diagonal $s\ell(2|1)$ in $2s\ell(2|1)$.

We use the notation $e_{i,j}$ for the matrix basis $(e_{i,j})_{kl}=
\delta_{ik}\delta_{jl}$. In the
fundamental representation, an element of $s\ell(4|3)$ will be represented by
an $7\times7$
supertrace-less matrix.
We define as first $s\ell(2|1)$ subalgebra
\be
\begin{array}{l}
F^{(1)}_{++}=e_{5,2} \mb{;} F^{(1)}_{+-}=e_{1,5} \mb{;} F^{(1)}_{-+}=e_{5,1}
\mb{;}
F^{(1)}_{--}=e_{2,5} \\ \\
H^{(1)}=\half(e_{1,1}-e_{2,2}) \mb{;} E^{(1)}_+=e_{1,2} \mb{;} E^{(1)}_-=
e_{2,1}\\ \\
Y^{(1)}= \half(e_{1,1}+e_{2,2}+ 2 e_{5,5})
\end{array}\label{sl.1}
\ee
and for second algebra we take:
\be
\begin{array}{l}
F^{(2)}_{++}=e_{7,4} \mb{;} F^{(2)}_{+-}=e_{3,7} \mb{;} F^{(2)}_{-+}=e_{7,3}
\mb{;}
F^{(2)}_{--}=e_{4,7} \\ \\
H^{(2)}=\half(e_{3,3}-e_{4,4}) \mb{;} E^{(2)}_+=e_{3,4} \mb{;} E^{(2)}_-=
e_{4,3}\\ \\
Y^{(2)}= \half(e_{3,3}+e_{4,4}+ 2 e_{7,7})
\end{array}\label{sl.2}
\ee
Then, the generators of the diagonal $s\ell(2|1)$ superalgebra are defined
by\footnote{The minus sign is for latter
convenience}
$F=F^{(1)}-F^{(2)}$ and $B=B^{(1)}+B^{(2)}$, where $F^{(i)}\ (B^{(i)}),\ i=1,2$
are the
fermionic (bosonic) generators of
the two $s\ell(2|1)$ (\ref{sl.1}) and (\ref{sl.2}).
Besides the two highest weights built on the bosonic roots of the $s\ell(1|2)$
subalgebras
$E_+(0,1)=e_{1,2}+e_{3,4}$ and $W_1(0,1)=e_{3,4}-e_{1,2}$, the highest weights
for the
diagonal $s\ell(2|1)$ are

\be \begin{array}{l}
W_2(0,1)= e_{1,4}\mb{;} W_3(0,1)= e_{3,2}\\
 \\
W_1(\half,\half)= e_{3,6}\mb{;} W_2(\half,\half)= e_{1,6}\\
 \\
W_1(-\half,\half)= e_{6,2}\mb{;} W_2(-\half,\half)= e_{6,4}\\
 \\
W_1(0,0)= e_{2,4}+ e_{1,3}- e_{5,7}\mb{;}
W_2(0,0)= e_{3,1}+ e_{4,2}- e_{7,5}\\
 \\
W_3(0,0)= e_{1,1}+ e_{2,2}+ e_{5,5}+ e_{6,6}\mb{;}\\ \\
W_4(0,0)= e_{3,3}+ e_{4,4}+ e_{6,6}+ e_{7,7}\mb{;}
\end{array}
\ee
We have quoted in parenthesis the eigenvalues $(b,j)$ of the generator
w.r.t. $Y$ and $H$ respectively. This corresponds to the decomposition
$4\ (0,1)\oplus\ 2\ (\half,\half)\oplus\ 2\ (-\half,\half)\oplus 4\ (0,0)$. It
is easily obtained using the
 decomposition of the fundamental obtained from two regular
$s\ell(2|1)$-subalgebras in $s\ell(4|3)$:
$\und{7}= 2(\half,\half)\oplus (0,0)$ and computing $7\times\bar{7}-\und{1}$.

Now, instead of starting from (\ref{sl.2}) as second $s\ell(2|1)$, we could
have chosen:
\be
\begin{array}{l}
F^{(2)}_{+-}=e_{7,4} \mb{;} F^{(2)}_{++}=e_{3,7} \mb{;} F^{(2)}_{--}=-e_{7,3}
\mb{;}
F^{(2)}_{-+}=-e_{4,7} \\ \\
H^{(2)}=\half(e_{3,3}-e_{4,4}) \mb{;} E^{(2)}_+=e_{3,4} \mb{;} E^{(2)}_-=
e_{4,3}\\ \\
Y^{(2)}= -\half(e_{3,3}+e_{4,4}+ 2 e_{7,7}) \end{array}\label{sl.2'}
\ee
Comparing (\ref{sl.2}) with (\ref{sl.2'}), it is clear that the two
$s\ell(2|1)$ are
isomorphic.
However, apart from $E_+(0,1)=e_{1,2}+e_{3,4}$ and $W(0,1)=e_{3,4}-e_{1,2}$,
the highest
weights of the diagonal $s\ell(2|1)$ become

 \be
\begin{array}{l}
W(1,1)= e_{1,4}\mb{;} W(-1,1)= e_{3,2}\\ \\
W_1(\half,\half)= e_{6,4}\mb{;} W_2(\half,\half)= e_{1,6}\\ \\
W_1(-\half,\half)= e_{6,2}\mb{;} W_2(-\half,\half)= e_{3,6}\\ \\
W(\frac{3}{2},\half)= e_{5,4}+ e_{1,7}\mb{;} W(-\frac{3}{2},\half)= e_{7,2}+
e_{3,5}\\ \\
W_1(0,0)= e_{1,1}+ e_{2,2}+ e_{5,5}+ e_{6,6}\\ \\
W_2(0,0)= e_{3,3}+ e_{4,4}+ e_{6,6}+ e_{7,7}\mb{;}
\end{array}
\ee
which gives the decomposition $2\ (0,1)\oplus\ (1,1)\oplus\ (-1,1)\oplus\
(\frac{3}{2},\half)
\oplus\ (-\frac{3}{2},\half)\oplus\ 2\ (\half,\half)\oplus\ 2\
(-\half,\half)\oplus 2\ (0,0)$.
As announced, this decomposition is obtained from the fundamental
$\und{7}= (\half,\half)\oplus (-\half,\half)\oplus (0,0)^\pi$.

Thus, we see that starting from $2s\ell(2|1)$ in $s\ell(4|3)$, one can take as
decomposition of the fundamental either $\und{7}= 2(\half,\half)\oplus (0,0)$,
or $\und{7}= (\half,\half)\oplus (-\half,\half)\oplus (0,0)^\pi$, depending on
the normalization one chooses. The corresponding adjoint decomposition will be
different but equivalent to the first one.

\indent

To conclude, let us add that doing the folding of $s\ell(4|3)$ to
get\footnote{Be carefull that the
$sp(4)$ subalgebra is in the upper left block instead of lower right block}
$osp(3|4)$, we have to
apply the rules
\be
\begin{array}{l}
e_{i,j}\equiv (-)^{i+j+1}\ e_{5-j,5-i} \mb{for} i,j=1,2,3,4\\ \\ e_{i,j}\equiv
(-)^{i+j+1}\ e_{12-j,12-i}
\mb{for} i,j=5,6,7\\ \\ e_{i,j}\equiv (-)^{i+j}\ e_{12-j,5-i} \mb{for}
i=1,2,3,4;\ j=5,6,7
\end{array}
\ee
These rules are clearly incompatible with the first choice of normalisation
(\ref{sl.2}), since they
impose $Y=Y^{(1)}+Y^{(2)}\equiv0$.
On the contrary, the
second choice of normalisation (\ref{sl.2'}) survive the folding procedure.
Thus,
the decomposition of the $osp(3|4)$ fundamental must be
$\und{7}= (\half,\half)\oplus (-\half,\half)\oplus (0,0)^\pi$. This is
an illustration of the "sign fixing" of atypical representations that occurs in
the $osp(m|2n)$ superalgebras.
Moreover, looking at the $s\ell(2|1)$-representation, one realises that we have
the identities (written below for the highest weights, but true for any
generator of the corresponding
$s\ell(2|1)$-representation):
\be
\begin{array}{l}
W(0,1)\equiv0\mb{;} W(\frac{3}{2},\half)\equiv0\mb{;}
W(-\frac{3}{2},\half)\equiv0\\ \\
W_1(\half,\half)\equiv W_2(\half,\half)\mb{;} W_1(-\half,\half)\equiv
W_2(-\half,\half)\mb{;}
W_1(0,0)\equiv W_2(0,0)
\end{array}
\ee
{}From these identities, one deduces that $osp(3|4)$ decomposes as
$(0,1)\oplus\ (1,1)\oplus\
(-1,1)\oplus\ (\half,\half)\oplus\ (-\half,\half)\oplus 2\ (0,0)$. Once more,
this result is easily
obtained from the product
\[
\Sym{\r{\half}\oplus \rmo{\half}}^2\oplus \Asym{(0,0)}^2\oplus
\left[\left(\shalf,\shalf) \oplus (-\shalf,\shalf\right)\right] \times (0,0)
\]

\begin{table}[p]
\begin{center}
\begin{tabular}{|c|c|c|c|} \hline
$\cg$ &SSA in $\cg$ &Fundamental of $\cg$ & Adjoint of \cg\\ \hline
&&& \\
$s\ell(1|2)$
&$s\ell(1|2)$ &$\rpi{\half}$ & $\ro{1}$ \\ &&& \\ \hline &&& \\
$s\ell(1|3)$
&$s\ell(1|2)$ &$\rpi{\half} \oplus \rpi{0}$ & $\ro{1}\oplus\r{\half}'
\oplus\rmo{\half}'\oplus\ro{0}$ \\ &&& \\ \hline &&& \\
$s\ell(2|2)$
&$s\ell(1|2)$ &$\rpi{\half} \oplus \r{0}$ & $\ro{1}\oplus \r{\half}
\oplus\rmo{\half}$\\
&&& \\ \hline &&& \\
$s\ell(1|4)$
&$s\ell(1|2)$ &$\rpi{\half} \oplus 2\rpi{0}$ & $\ro{1}\oplus 2\r{\half}'
\oplus2\rmo{\half}' \oplus4\ro{0}$\\ &&& \\ \hline &&& \\
$s\ell(2|3)$
&$s\ell(2|3)$ &$\rpi{1}$ & $\ro{2}\oplus\ro{1}$\\ &&& \\
&$s\ell(1|2)$
&$ \rpi{\half} \oplus \r{0} \oplus \rpi{0}$ & $\begin{array}{c}
\ro{1}\oplus \r{\half}\oplus\rmo{\half}\oplus\\
 \oplus\r{\half}'\oplus\rmo{\half}' \oplus2\ro{0} \oplus2\ro{0}' \end{array}$
\\
&&& \\
&$s\ell(2|1)$ &$\r{\half} \oplus 2\rpi{0}$ & $\ro{1} \oplus 2\r{\half} \oplus
2\rmo{\half} \oplus 4\ro{0}$\\ &&& \\ \hline
\end{tabular}
\caption{ $s\ell(m|n)$ superalgebras up to rank 4. \label{T1}} \end{center}
\end{table}

\begin{table}[p]
\begin{center}
\begin{tabular}{|c|c|c|c|} \hline
$\cg$ &SSA in $\cg$ &Fundamental of $\cg$ & Adjoint of \cg\\ \hline
&&& \\
$osp(1|2n) $ & $\emptyset$ & $-$ & $-$ \\
&&& \\ \hline &&& \\
$osp(3|2)$
& $ osp(2|2) $ &$\ropi{\half} \oplus \r{0}$ & $\ro{1}\oplus \ro{\half}$ \\
&&& \\ \hline &&& \\
$osp(3|4)$
& $osp(2|2)$ & $ \ropi{\half} \oplus \r{0} \oplus 2\rpi{0} $ &
$\begin{array}{c}
 \ro{1}\oplus \ro{\half}\oplus 2\ro{\half}' \oplus\\ \oplus 3\ro{0} \oplus
2\ro{0}'
\end{array}$\\
&&& \\
& $ sl(1|2) $ &$\rpi{\half}\oplus\rpim{\half} \oplus \r{0} $ &
$\begin{array}{c}
\ro{1}\oplus \r{1} \oplus\rmo{1} \oplus\\ \oplus\r{\half}\oplus \rmo{\half}
\oplus \ro{0} \end{array}$\\
&&& \\ \hline &&& \\
$osp(5|2)$
&$ osp(2|2) $
&$\ropi{\half} \oplus 3\r{0}$ & $\ro{1}\oplus 3\ro{\half} \oplus3\ro{0}$\\
&&& \\
&$s\ell(2|1)$ &$\r{\half}\oplus\rmo{\half} \oplus \r{0}$ & $\begin{array}{c}
\ro{1}\oplus \rge{\frac{3}{2}}{\half} \oplus
\rge{\mbox{-}\frac{3}{2}}{\half}\oplus\\ \oplus\r{\half}'\oplus\rmo{\half}'
\oplus\ro{0} \end{array}$ \\
&&& \\ \hline
\end{tabular}
\caption{ $osp(2m+1|2n)$ superalgebras of rank $<$4.\label{T2}} \end{center}
\end{table}

\begin{table}[p]
\begin{center}
\begin{tabular}{|c|c|c|c|} \hline
$\cg$ &SSA in $\cg$ &Fundamental of $\cg$ & Adjoint of \cg\\ \hline
&&& \\
$osp(3|6)$
&$ s\ell(1|2) $
&$\begin{array}{c}
\rpi{\half}\oplus \rpim{\half}\oplus \\ \oplus \r{0} \oplus 2\rpi{0}
\end{array}$
& $ \begin{array}{c} \ro{1} \oplus \r{1}\oplus \rmo{1}\oplus\\
\oplus\r{\half}\oplus\rmo{\half} \oplus2\r{\half}'\oplus \\ \oplus
2\rmo{\half}' \oplus4\ro{0}\oplus 2\ro{0}' \end{array}$\\
&&& \\
&$ osp(2|2) $
&$ \ropi{\half} \oplus 4\rpi{0} \oplus \r{0}$ &$\begin{array}{c} \ro{1} \oplus
\ro{\half} \oplus 4\ro{\half}' \oplus \\ \oplus 10\ro{0}\oplus 4\ro{0}'
\end{array}$\\
&&& \\ \hline &&& \\
$osp(5|4)$
&$2\ osp(2|2)$ & $2\ropi{\half}\oplus (0,0)$ &$3\ro{1}\oplus 2\ro{\half} \oplus
[(0,\half)]^2_A$ \\
&&& \\
&$s\ell(1|2) $& $ \rpi{\half}\oplus\rpim{\half} \oplus 3\r{0}$
&$\begin{array}{c} \ro{1}\oplus\r{1}\oplus\rmo{1}\oplus\\ \oplus
3\r{\half}'\oplus 3\rmo{\half}'\oplus 4\ro{0}\end{array}$ \\
&&& \\
&$ osp(2|2)$ &$\ropi{\half} \oplus 3\r{0} \oplus 2\rpi{0}$ & $\begin{array}{c}
\ro{1}\oplus 3\ro{\half}\oplus 2\ro{\half}'\oplus\\ \oplus6\ro{0}\oplus
6\ro{0}' \end{array}$\\
&&& \\
&$s\ell(2|1)$ &$\begin{array}{c} \r{\half}\oplus \rmo{\half} \oplus \\
\r{0} \oplus 2\rpi{0}\end{array}$
& $\begin{array}{c}
\ro{1}\oplus \rge{\frac{3}{2}}{\half} \oplus
\rge{\mbox{-}\frac{3}{2}}{\half}\oplus \\ \oplus 2\r{\half}\oplus 2\rmo{\half}
\oplus \\ \oplus \r{\half}' \oplus\rmo{\half}' \oplus 4\ro{0} \oplus 2\ro{0}'
\end{array}$\\
&&& \\ \hline &&& \\
$osp(7|2)$
&$ osp(2|2)$ &$\ropi{\half} \oplus 5\r{0}$ & $\ro{1}\oplus 5\ro{\half} \oplus
10\ro{0}$ \\ &&& \\
&$s\ell(2|1)$ &$\r{\half}\oplus\rpim{\half} \oplus 3\r{0}$ & $\begin{array}{c}
\ro{1}\oplus \rge{\frac{3}{2}}{\half} \oplus
\rge{\mbox{-}\frac{3}{2}}{\half}\oplus \\ \oplus 3\r{\half}' \oplus
3\rmo{\half}' \oplus4\ro{0} \end{array}$\\
&&& \\ \hline
\end{tabular}
\caption{ $osp(2m+1|2n)$ superalgebras of rank 4.\label{T2bis}} \end{center}
\end{table}

\begin{table}[p]
\begin{center}
\begin{tabular}{|c|c|c|c|} \hline
$\cg$ &SSA in $\cg$ &Fundamental of $\cg$ & Adjoint of \cg\\ \hline
&&& \\
$osp(4|2)$
&$osp(2|2)$ &$\ropi{\half} \oplus 2\r{0}$ &$\ro{1}\oplus 2\ro{\half}\oplus
\ro{0}$ \\ &&& \\
&$s\ell(2|1)$ &$\r{\half}\oplus \rmo{\half}$ & $\ro{1}\oplus
\rge{\frac{3}{2}}{\half}\oplus \rge{\mbox{-}\frac{3}{2}}{\half} \oplus
\ro{0}$\\
&&& \\ \hline &&& \\
$osp(4|4)$
&$osp(2|2)$ &$\ropi{\half} \oplus 2\r{0} \oplus 2\rpi{0}$ & $\begin{array}{c}
\ro{1}\oplus 2\ro{\half} \oplus 2\ro{\half}' \\ \oplus 4\ro{0}\oplus 4\ro{0}'
\end{array}$\\
&&& \\
&$ 2\ osp(2|2) $
&$2\ropi{\half}$ & $
3\ro{1}\oplus \rinde $\\
&&& \\
& $s\ell(1|2) $ & $\rpi{\half}\oplus\rpim{\half} \oplus 2(0,0)$
&$\begin{array}{c} \ro{1}\oplus \r{1}\oplus \rmo{1} \oplus\\ \oplus
2\r{\half}\oplus 2\rmo{\half} \oplus 2\ro{0} \end{array}$\\
&&& \\
&$s\ell(2|1)$ &$\r{\half}\oplus\rmo{\half} \oplus 2\rpi{0}$ & $\begin{array}{c}
\ro{1}\oplus \rge{\frac{3}{2}}{\half}\oplus
\rge{\mbox{-}\frac{3}{2}}{\half}\oplus \\ \oplus 2\r{\half} \oplus 2\rmo{\half}
\oplus4 \ro{0} \end{array}$ \\
&&& \\ \hline &&& \\
$osp(6|2)$
&$osp(2|2)$ &$\ropi{\half} \oplus 4\r{0}$ &$\ro{1}\oplus 4\ro{\half}\oplus
6\ro{0}$ \\
&&& \\
&$s\ell(2|1)$ &$\r{\half}\oplus \rmo{\half} \oplus 2\r{0}$ &$\begin{array}{c}
\ro{1}\oplus \rge{\frac{3}{2}}{\half} \oplus \rge{\mbox{-}\frac{3}{2}}{\half}
\oplus\\ \oplus 2\r{\half}'\oplus 2\rmo{\half}' \oplus 2\ro{0} \end{array}$ \\
&&& \\ \hline
&&& \\
$osp(2|4)$
&$s\ell(1|2)$ &$\rpi{\half}\oplus\rpim{\half}$ & $\ro{1}\oplus \r{1}\oplus
\rmo{1}\oplus \ro{0} $\\ &&& \\
&$osp(2|2)$ &$\ropi{\half} \oplus 2\rpi{0}$ & $\ro{1}\oplus 2\ro{\half}'\oplus
3\ro{0}$ \\ &&& \\ \hline &&& \\
$osp(2|6)$
&$s\ell(1|2)$
&$\rpi{\half}\oplus \rpim{\half} \oplus 2\rpi{0}$ &$\begin{array}{c}
\ro{1}\oplus \r{1}\oplus \rmo{1}\oplus \\ \oplus 2\r{\half}'\oplus
2\rmo{\half}'\oplus4 \ro{0} \end{array}$ \\
&&& \\
&$osp(2|2)$ &$\ropi{\half} \oplus 4\rpi{0}$ & $\ro{1}\oplus 4\ro{\half}'\oplus
10\ro{0}$\\ &&& \\ \hline
\end{tabular}
\caption{ $osp(2m|2n)$ superalgebras up to rank 4.\label{T3}} \end{center}
\end{table}

\begin{table}[p]
\begin{center}
\begin{tabular}{|c|c|c|} \hline
\cg &SSA in\cg& $s\ell(1|2)$ decomposition \\ \hline
&& \\
$G(3)$& $s\ell(2|1)$ &
$(0,1)\oplus (\frac{5}{6},\half)\oplus (\mbox{-}\frac{5}{6},\half)
\oplus (\frac{1}{6},\half)'\oplus (\mbox{-}\frac{1}{6},\half)'
\oplus (\frac{1}{2},\half)'\oplus (\mbox{-}\frac{1}{2},\half)'\oplus (0,0)
$ \\
& &\\
& $s\ell(2|1)'$ &
$(0,1)\oplus (\frac{7}{2},\half)\oplus (\mbox{-}\frac{7}{2},\half)
\oplus (\frac{3}{2},\frac{3}{2})'\oplus
(\mbox{-}\frac{3}{2},\frac{3}{2})'\oplus
 (0,0)$ \\
& &\\
& $osp(2|2)$ &
$(0,1)\oplus 2\, (\frac{1}{4},\half)\oplus 2\, (\mbox{-}\frac{1}{4},\half)
\oplus (0,\frac{1}{2})$ \\
& &\\ \hline
& &\\
$F(4)$& $s\ell(2|1)$ & $(0,1)\oplus 3 (\frac{1}{6},\half)\oplus
3 (\mbox{-}\frac{1}{6},\half)\oplus 8 (0,0)$\\
& &\\
& $s\ell(1|2)$ & $(0,1)\oplus (1,\half)\oplus (\mbox{-}1,\half)\oplus 4 (0,0)
\oplus 2 (\half,\half)'\oplus 2 (\mbox{-}\half,\half)'\oplus 2 (0,\half)'$
\\
& &\\
& $osp(2|2)$ & $(0,1)\oplus 2 (1,1)\oplus 2 (\mbox{-}1,1)\oplus
(\frac{5}{2},\half)\oplus
(\mbox{-}\frac{5}{2},\half)\oplus 4 (0,0)$\\ & &\\ \hline
& &\\
$D(2,1;\alpha)$& $osp(2|2)$ & $\ro{1}\oplus 2\ro{\half}\oplus \ro{0}$ \\
& &\\
& $s\ell(2|1)$ & $\ro{1}\oplus \rge{\frac{3}{2}}{\half}\oplus
\rge{\mbox{-}\frac{3}{2}}{\half} \oplus\ro{0}$\\
 & &\\ \hline
\end{tabular}
\caption{ The exceptional superalgebras \label{T4}} \end{center}
\end{table}

\newpage

\sect{The standard reduction \label{standard}}
Consider an embedding of $sl(2|1)$ in some super Lie algebra \cg. We normalize
the $sl(2)$ subalgebra of $sl(2|1)$ such that $[e_0,e_\pp\, ]=+2e_\pp$,
$[e_0,e_=]=-2e_=$ and $[e_\pp\, ,e_=]=e_0$. The standard grading is
nothing but the $sl(2)$ grading, {\it i.e.} given by $ \frac 1 2
\mbox{ad}_{e_0}$. The action
\bea
{\cal S}_0=\k S^-[g]+\frac{1}{\pi x}\int str\, A(J-\frac \k 2 e_=-\frac \k 2
[e_=,{\tau}])
-\frac{\k}{4\p x}\int str [e_=,{\tau}] \bdel{\tau},\label{actionst}
\eea
where
\bea
A&\in&\Pi_{>0}\cg ,\nonu
{\tau}&\in&\Pi_{\frac 1 2}\cg
\eea
has a gauge invariance:
\bea
\d g&=& \h g,\nonu
\d A&=& \bdel \h + {[}\h, A{]},\nonu
\d {\tau}&=& - \Pi_{\frac 1 2}\h,\label{trsfst}
\eea
and $\h\in \Pi_{>0}\cg$. Gauge invariance requires the introduction of the
$\tau$ field\footnote{In
a Hamiltonian treatment, they are needed to obtain first class constraints.}.
The constraints and the resulting gauge invariance breaks the
original chiral affine symmetry of the WZW model to some extension of the $N=2$
superconformal algebra.
The generators are precisely the gauge invariant polynomials in $\P_{\geq 0}J$
and $\t$ and their
derivatives.
In order to quantize the model, we take $A=0$ as a gauge choice. Upon the
introduction of
the ghosts $c\in\P_{>0}\cg$ and anti-ghosts $b\in\P_{<0}\cg$, we get the gauge
fixed action:
\bea
{\cal S}_{\rm gf}=\k S^-[g]
+ \frac{\k}{4\p x}\int str [\t,e_=]\bdel\t
+\frac{1}{2\p x}\int str\, b\bdel c,
\label{actiongf}
\eea
and the BRST charge $\cq_{HR}$:
\bea
\cq_{HR}=\frac{1}{4\p i x}\oint str\left\{ c \left( J -\frac \k 2
e_=-\frac \k 2[e_=,\t]+ \frac 1 2 J^{\rm gh}\right) \right\}.\label{brsch1}
\eea
where $J^{\rm gh}=\frac 1 2 \{b,c\}$. The generators of the extended $N=2$
superconformal
algebra are now the generators of the cohomology of
$\cq_{HR}$ computed on the algebra $\ca$ which is generated by
$\{b,\hat{J},\t,c\}$, with $\hat{J}=J+J^{\rm
gh}$, and it
consists of all regularized
products of the generating fields and their
derivatives modulo the usual relations
between different orderings, derivatives, etc. The calculation of this
cohomology proceeds along exactly the
same lines as in \cite{ST}. To make this paper selfcontained, we summarize the
results.
The subcomplex $\ca^{(1)}$ generated by $\{b,\P_{<0}\hat{J}\}$ has a trivial
cohomology
$H^*(\ca^{(1)};\cq)={\bf C}$. The only field with negative ghost number is $b$.
This results in
that the full cohomology $H^*(\ca;\cq_{HR})$ is equal to the
cohomology $H^*(\wha;\cq_{HR})$, where we introduced
the reduced complex $\wha$ generated by $\{\P_{\geq 0}
\hat{J},\t,c\}$. The OPE's close on $\wha$. The underlying
$sl(2)$ grading implies the existence of a double grading on $\wha$:
\be
\wha=\bigoplus_{\stackrel{\scriptstyle m,n\in\frac 1 2{\bf Z}}{m+n\in{\bf
Z}}}\wha_{(m,n)},
\ee
where
\bea
X\in \wha_{(m,n)}\Longleftrightarrow {[}e_0,X{]}=2m \, X \mbox{ and }m+n = \
ghost number(X).
\eea
We assign to ${\tau}$ grading $(0,0)$. The BRST operator itself decomposes into
three parts, each of
definite grading: $\cq_{HR}=\cq_{(1,0)}+\cq_{(1/2,1/2)}+\cq_{(0,1)}$ with
\bea
\cq_{(1,0)}&=&-\frac{\k}{8\p i x}\oint str c e_=\nonu
\cq_{(1/2,1/2)}&=&-\frac{\k}{8\p i x}\oint str
c\left[{e_=,\tau}\right],\label{qsqr}
\eea
and from $\cq_{HR}^2=0$ one gets immediately
\bea
\cq_{(1,0)}^2 = \cq_{(0,1)}^2 = \{\cq_{(1,0)},
\cq_{(1/2,1/2)}\} =
\{\cq_{(0,1)}, \cq_{(1/2,1/2)}\} = \cq_{(1/2,1/2)}^2 + \{\cq_{(0,1)},
\cq_{(1,0)}\} =0.
\eea
The filtration
$\wha^m$, $m\in\frac 1 2 {\bf Z}$ of $\wha$:
\be
\wha^m\equiv \bigoplus_{k\in\frac 1 2 {\bf Z}}\bigoplus_{l\geq m}\wha_{(k,l)}.
\ee
leads to  a spectral sequence $(E_r,d_r)$, $r\geq1$, converging
to $H^*(\wha;\cq)$.  Each term in the sequence is given by the cohomology of
the previous
term with a derivation that represents the effective action of the
BRST operator at that level: $E_r=H^*(E_{r-1};d_{r-1})$.  The first term in the
sequence is then $E_0=\wha$, $d_0=\cq_{(1,0)}$. One readily computes $E_1$:
\bea
E_1\simeq\wha \left[\P_{\ker ad e_\pp}\hat{J} \right]\otimes\wha \left[
{\tau}\right]\otimes\wha \left[ \P_{\frac 1 2 }c\right],
\eea
the subsequent term has $d_1=\cq_{(1/2,1/2)}$ and one gets:
\bea
E_2\simeq\wha \left[\P_{\ker ad e_+}\left(\hat{J} +\frac\k
4[{\tau},[e_-,{\tau}]]\right)\right].
\eea
After this the spectral sequence collapses, and we get
\bea
H^*(\ca;\cq_{HR})\simeq E_2=H^*(H^*(\wha,\cq_{(1,0)}),\cq_{(1/2,1/2)}).
\eea
Of course this is only an isomorphism of the cohomologies as vectorspaces. In
order to get the generators of
$H^*(\wha;\cq_{HR})$ we use a generalized tic-tac-toe construction, which
determines the currents up to a scale
factor. Indeed if $X_{(j,-j)}$ belongs to  $\hat{J} +\frac\k
4[{\tau},[e_-,{\tau}]]$ and has fixed grading $(j,-j)$ we obtain the full
generator $\hat{X}_j$
\bea
\hat{X}_j=\sum_{2m=0}^j X_{(m,-m)},
\eea
where $X_{(m,-m)}$ are recursively determined from
\bea
\{\cq_{(0,1)},X_{(n,-n)}\}+
\{\cq_{(1/2,1/2)},X_{(n-1/2,-n+1/2)}\}+
\{\cq_{(1,0)},X_{(n-1,-n+1)}\}=0.
\eea
In this way we get as many BRST invariant currents as there are $sl(2)$ irreps
present in the decompostion of
the adjoint representation. The OPE's of these currents close: by construction,
the OPE of two generators of
$H^*(\wha;\cq_{HR})$ closes modulo BRST exact terms. However, as we found that
the cohomology is only
non-trivial in the ghost number zero sector and as we computed our cohomology
on the reduced
complex $\wha$ which has no negative ghost number currents we get that the
OPE's of the generators of
$H^*(\wha;\cq_{HR})$ close.
The quantum Miura transformation follows for free from this construction.
The map $\hat{X}_j\rightarrow X_{(0,0)}$ is obviously an algebra homomorphism.
In fact it is also an algebra
isomorphism. To see this, one only has to show that for each $\hat{X}_j$,
$X_{(0,0)}$ is non-vanishing.
Consider for this the mirror
of the spectral sequence, {\it i.e.} the one which follows from the filtration
\be
\wha'^m\equiv \bigoplus_{l\in\frac 1 2 {\bf Z}}\bigoplus_{k\geq m}\wha_{(k,l)}.
\ee
A small computation teaches us that $E_1=H^*(\wha ;
Q_{(0,1)})$ is non-vanishing only for grades $(\frac m 2 ,\frac m 2)$, $m\geq
0$. We already know that
$E_\infty$ is non trivial only at ghost number 0. Combining these statements
shows that $X_{(0,0)}$ is always
non-trivial.
Till now we only used the underlying $sl(2)$ embedding in $\cg$. However the
full $sl(2|1)$ embedding is
relevant
in that it guarantees that the resulting algebra is an extension of the $N=2$
superconformal algebra. So what
remains to be shown is that the resulting conformal algebra has an $N=2$
superconformal subalgebra. That this
is the case follows immediately from the lemma \cite{jdb}:
\begin{lem}\label{lemn}
If the conformal algebra $\cv_1$ is obtained from the Hamiltonian reduction
based on the algebra homomorphism
$i_1$: $sl(2)\rightarrow \cg_1$ and $\cv_2$ from $i_2$: $sl(2)\rightarrow
\cg_2$, then $\cv_1\subseteq \cv_2$
if any of the following two cases are satisfied:
\begin{enumerate}
\item $\cg_2=\cg_1\oplus\cg '$ and $i_2\mid_{\cg_1}=i_1$.
\item There is an algebra homomorphism $j: \cg_1\rightarrow\cg_2$, such that
$i_2=j \circ i_1$.
\end{enumerate}
\end{lem}
Obviously 2. is satisfied here. Rests us to determine the central extension of
the algebra. One takes the
energy-momentum tensor, improved by a BRST exact piece such as to get it in a
Sugawara like form:
\bea
\hat{T}^{\rm IMP}
&\equiv&\frac{1}{x(\k+\tilde{h})} str J J  -\frac{1}{8xy}str e_0 \del J
-\frac{\k}{4 x}str\left( [\t,e_=]\del\t\right)\nonu
&&+\frac{1}{4x}str b[e_0,\del c]-\frac {1}{2x}str b\del c+\frac{1}{4x}str
\del b
[e_0,c]
,\label{Timp}
\eea
where $y$ is the index of embedding and one finds
\bea
c=\frac 1 2 c_{\rm crit} - \frac{(d_B-d_F)\tilde{h}}{\k+\tilde{h}} - 6 y (\k
+\tilde{h}),\label{cpretty}
\eea
where $c_{\rm crit}$ is the expected critical charge of the string, {\it i.e.}
the value of $c$ for which the conformal anomaly cancels:
\bea
c_{\rm crit}=\sum_{j,\a_j}(-)^{(\a_j)}(12 j^2+12j+2),\label{ccrit}
\eea
where the sum runs over all $sl(2)$ representations $\underline{2j+1}$
occuring in the decomposition of the adjoint representation of
$\cg$, $\a_j$ is the multiplicity of a
representation $\underline{2j+1}$ and the phase $(-)^{(\a_j)}$ is $+1$, $-1$
resp., if the representation has a bosonic, fermionic resp., nature. Finally
$y$ is the index of embedding. A particularly useful expression for it is given
by
\bea
y=\frac{1}{3\tilde{h}}\sum_{j,\a_j} (-)^{(\a_j)}j(j+1)(2j+1).\label{indexem }
\eea
Summarizing, we arrive at the following picture. One starts with an embedding
of $sl(2|1)$ in  some super Lie
algebra $\cg$. The adjoint representation of $\cg$ decomposes into irreps of
$sl(2|1)$ as
\bea
\mbox{adjoint}(\cg)=\bigoplus_{j\in\frac 1 2 {\bf N},b\in\frac 1 2 {\bf
Z}}n_{(b,j)}\, (b,j),\label{decomp}
\eea
with $n_{(b,j)}\in{\bf N}$, the multiplicities. Performing the reduction in the
way we discussed above yields
an extension of the $N=2$ superconformal algebra with central charge given in
eq. (\ref{cpretty}). The
embedded $sl(2|1)$ gives then rise to the $N=2$ superconformal algebra. For the
remainder we get that every
$sl(2|1)$ irrep $(b,j)$, $b\neq \pm j$, gives rise to a set of 4 conformal
currents $(Z,H_+,H_-,Y)$ which form a primary,
unconstrained $N=2$ multiplet of conformal dimension $j$ and charge $2b$.
The OPE's of the untwisted $N=2$ generators $T_{N=2},\ G_\pm$ and $U$ with
these currents follow from $N=2$ representation theory:
\bea
&&T_{N=2}(z_1)Z(z_2)=hz_{12}^{-2} Z(z_2)+
z_{12}^{-1}\del Z(z_2),\nonu
&&T_{N=2}(z_1)H_\pm(z_2)=(h+\frac 1 2)z_{12}^{-2}H_\pm(z_2)+z_{12}^{-1}\del
H_\pm(z_2),\nonu
&&T_{N=2}(z_1)Y(z_2)=\frac q 2 z_{12}^{-3}Z(z_2)+
(h+1)z_{12}^{-2}Y(z_2)+z_{12}^{-1}\del Y(z_2),\nonu
&&G_\pm(z_1)Y(z_2)= (h\pm\frac q 2+\frac 1 2 )
z_{12}^{-2} H_\pm(z_2)+\frac 1 2 z_{12}^{-1}\del H_\pm (z_2),\nonu
&&G_+(z_1)H_-(z_2)= (h+\frac q 2)z_{12}^{-2}Z(z_2) +
z_{12}^{-1} (Y(z_2)+\frac 1 2 \del Z(z_2)),\nonu
&&G_-(z_1)H_+(z_2)= -(h-\frac q 2)z_{12}^{-2}Z(z_2) +
z_{12}^{-1} (Y(z_2)-\frac 1 2 \del Z(z_2)),\nonu
&&G_\pm(z_1)Z(z_2)= \mp
z_{12}^{-1} H_\pm(z_2),\nonu
&&U(z_1)Z (z_2)=q z_{12}^{-1}Z (z_2), \qquad
U(z_1)H_\pm (z_2)=(q\pm 1)z_{12}^{-1}H_\pm (z_2), \nonu
&&U(z_1)Y(z_2)=hz_{12}^{-2}Z(z_2)+q z_{12}^{-1}Y (z_2),
\label{twn4}
\eea
where $j=h$ and $q=2b$.

For atypical
$sl(2|1)$ representations $(j, j)$ or $(-j, j)$, we get 2 conformal
currents which form a primary chiral or anti-chiral $N=2$ multiplet of
conformal dimension $j$ and charge $2j$ or $-2j$. {\it E.g.} for $(j, j)$,
we get currents $Z$ and $H_-$, whose OPE's follow from eq. (\ref{twn4}) by
putting $b=j$ (or $q=2h$) and setting $Y=\frac 1 2 \del Z$ and $H_+=0$.
Similar statements hold for
the $(-j,j)$ case, where $q=-2h$ and one puts $H_-=0$ and $Y=-\frac 1 2 \del
Z$.

One more subtlety has to be mentioned here. In the case that $j=0$ or
$j=1/2$, the conformal multiplets are not complete. Indeed in the first
case, 2 fields of conformal dimension $1/2$ and one of conformal dimension
$0$ (which only appears through its derivative in the algebra) are
lacking, while in the second case one scalar is missing. This is due to
the fact that neither scalars nor dimension $1/2$ fermions can be
generated through Hamiltonian reduction. By redoing the previous analysis
in $N=1$ superspace, see {\it e.g.} \cite{eric}, one does generate
dimension $1/2$ fermions, but the scalars are still missing. However, we
can repair this situation by reversing the Goddard-Schwimmer scheme,
\cite{goddard}. They showed that dimension $1/2$ and $0$ fields can always be
decoupled from the conformal algebra. This transformation turns out to be
invertible and so can be applied here.

Finally,
we did not take into account in
eq. (\ref{decomp}) that multiple $osp(2|2)$ embeddings, the adjoint is not
fully reducible in terms of
$sl(2|1)$ representations. Presumably, this will give rise to a new type of
$N=2$ representations. The study
of those will be reported on elsewhere.
\sect{The general construction \label{general}}
\subsection{Some general considerations}
We leave the standard reduction behind us and introduce a different grading
which will allow for a
stringy interpretation of the reduction. We
choose an embedding of $sl(2|1)$ in some super Lie algebra \cg. The adjoint
representation of $\cg$  decomposes
into a number of irreducible $sl(2|1)$ representations, see eq. (\ref{decomp}).
{}From
the example in section 2, we expect $2 \, b$ to be identified with the
ghost number of the resulting currents. So it is quite natural to
focus on the case where only $sl(2|1)$ representations with $b=0$ occur. One
finds that this happens for
\begin{enumerate}
\item $sl(m|n)$\\
A principal embedding of $sl(2|1)$ in $p\, sl(2j+1|2j)\oplus q\, sl(2j|2j+1)$,
which in its turn is regularly
embedded in $sl(m|n)$ with $p,\, q\geq 0$, $j\in\frac 1 2 {\bf N}$,
$m=p(2j+1)+2qj$ and $n=2pj+q(2j+1)$.
\item $osp(m|2n)$\\
The diagonal embedding of $osp(2|2)$ in $k\,(osp(2|2))$ which in its turn sits
regularly embedded in
$osp(m|2n)$. However, only for $k=1$ is the adjoint representation
fully reducible. So we only take $k=1$ into account.
\item $D(2,1,\a)$\\
$osp(2|2)$ as a regular subalgebra of $D(2,1,\a)$.
\end{enumerate}
To show this, one uses the results of section 4 from which it followed that the
$sl(2|1)$ decomposition of the
adjoint of $\cg$ follows from the products $(0,1/2)\otimes (0,1/2)$,
$(0,1/2)\otimes (\pm j,j)$,
$(j,j)\otimes (\pm k,k)$ and $(-j,j)\otimes (\pm k,k)$. Excluding the cases
where non fully reducible
representations occur, we find
\bea
(0,\frac 1 2)\otimes (0,\frac 1 2)\mid_{S}&=&(0,1)\nonu
(0,\frac 1 2)\otimes (\pm j,j)&=&(\pm j,j+\frac 1 2 )\oplus (\pm (j+\frac 1
2),j)\nonu
(j,j)\otimes (k,k)&=&
\bigoplus_{l=|j-k|+\frac 1 2 }^{j+k-\frac 1 2}(j+k+\frac 1 2 , l) \oplus
(j+k,j+k)\nonu
(j,j)\otimes (-k,k)&=&\bigoplus_{l=|j-k|}^{j+k}(j-k, l).
\eea
{}From this it follows that $b=0$ if only $(0,1/2)\otimes (0,1/2)$ and
$(j,j)\otimes (-j,j)$ occur.
For the fundamental representation of $sl(m|n)$ this implies that only the
regular embedding of $p\,
sl(2j+1|2j)\oplus q\, sl(2j|2j+1)$ should be considered for which the
fundamental representation decomposes as
\bea
\underline{m+n}=p\, (j,j)\oplus q\, (j,j)^\pi.
\eea
{}From the last equation we get that $m=p(2j+1)+q2j$ and $n=q(2j+1)+p2j$.
A similar analysis applies to the case of $osp(m|2n)$. The exceptional algebras
$\cg=G(3)$, $F(4)$ or
$D(2,1,\a)$ follow from direct inspection. $D(2,1,\a)$ is explicitely given in
table 5 and one gets that the
regular $osp(2|2)$ embedding leads to a decomposition with $b=0$. For both
$G(3)$ and $F(4)$, one finds always $sl(2|1)$ representations with
$b\neq 0$.
We take a
grading given by\footnote{Our normalizations are such that for a highest weight
state of an $sl(2|1)$ irrep,
$t_{(b,j)}$, ${[}e_0,t_{(b,j)}{]}=2j\, t_{(b,j)}$ and
${[}u_0,t_{(b,j)}{]}=2b\, t_{(b,j)}$.}
$ \frac 1 2 \mbox{ad}_{e_0}+ l\, \mbox{ad}_{u_0}$, where $l$ is some
positive
integer which will be determined now. An $sl(2|1)$ irrep $(b,j)$ decomposes
into $u(1)\oplus sl(2)$
irreps as $(b,j)=
|b,j>\oplus |b-\half,j-\half>\oplus |b+\half,j-\half>\oplus |b,j-1>$ where we
have $b=0$. After
the reduction we expect that for
each $sl(2)$ irrep, there will be one conformal current. The conformal current
associated to the $|0,j>$, with
conformal dimension $j+1$,
should be such that, after twisting, we can identify it with a total (so
including matter, ghost and gravity
contributions) symmetry current of the string theory.
Furthermore we want to identify the anti-ghost corresponding to this symmetry
with the current which is
associated to the $|-\half , j-\half >$ irrep. In order to achieve this we want
the highest weight of
$|-\half , j-\half >$ to be of negative grading so that its constraint puts it
equal to an auxiliary field.
We can achieve this by choosing $l$ in the grading as $l>j_{max}-\half$, where
$j_{max}$ is the largest value
of $j$ appearing in
the decomposition of the adjoint of \cg. For $osp(2|2) \hookrightarrow
osp(m|2n)$ one gets $l=1$ and for
$sl(2|1) \rightarrow p\, sl(2j+1|2j) \oplus q \, sl(2j|2j+1) \hookrightarrow
sl(p(2j+1)+2qj| 2pj +
q(2j+1))$ we get that $l=2j$.
For generic\footnote{We do the counting here for generic representations
$(0,j)$, the cases $(0,0)$ and $(0,\half)$ have to be done separately. We leave
this to the reader.} values of
$j$, we get that the $sl(2|1)$ irrep $(0,j)$ is $8j$ dimensional.
If $j$ in $(0,j)$ is an integer then one finds that
$4j-1$ elements of the representation have a strict positive grading. This
gives us both the dimension of the
gauge group and the number of constraints. Subtracting this from the original
number of affine currents, we
are left with 2 currents. However we still have to introduce auxiliary fields
of the $\t$ and the
$\psi,\ \bar{\psi}$ type. No fields of the $\t$ type are needed as
$\P_{-1/2}\P_{im\,ad\,e_{=}}\cg = \emptyset$.
We only
have one affine current which is both a highest $sl(2)$ weight and negatively
graded, so one set of
$\psi,\ \bar{\psi}$ fields has to be introduced. In total this leaves us 4
currents, the correct number of
degrees of freedom.
If $j$ is a half integer we get a slightly different counting. The gauge group
has now dimension $4j$. Again
we have to introduce one set of  $\psi,\ \bar{\psi}$ fields to account for the
negatively graded highest
$sl(2)$ weight state and we also have that $\P_{-1/2}\P_{im\,ad\,e_=}\cg$
contains two elements, requiring the
introduction of two $\t$ fields. Doing the counting gives us that again, as it
should be, 4 currents are left.
We thus come to the following picture. Performing the reduction with the above
grading will yield for each
$sl(2|1)$ irrep 4 conformal currents which form a standard $N=2$ unconstrained
multiplet. After twisting, we
identify the current associated with $|\frac 1 2,\frac 1 2>$ component of the
embedded $sl(2|1)$ itself with
the BRST current. From standard $N=2$ representation theory, it follows that
after twisting, each total
current of the string theory, associated with the $|0,j>$ component of $(0,j)$
is the BRST transform of the
corresponding anti-ghost, which is associated to the $|-\frac 1 2,j-\frac 1 2>$
component of $(0,j)$.
We briefly return to the general case where also representations with $b\neq 0$
occur.
One expects that the $b=0$ subsector will
have a direct ``stringy'' interpretation. However, in order that the BRST
structure closes one needs here the
introduction of extra $N=2$ multiplets with a non-vanishing ghost number. It
remains obscure how elementary
objects with ghost number greater than one can ever arise in a string theory.
So, for
the moment we do not discuss the
$b\neq 0$ case. We will come back to this case in a future publication.
\subsection{The invariant action}
{}From now on, we only use the grading discussed above and furthermore, we only
consider the cases where
$sl(2|1)$ representations occur which are fully reducible and which have $b=0$.
The action
\bea
{\cal S}_0=\k S^-[g]+\frac{1}{\pi x}\int str\, A(J-\frac \k 2 e_=-\frac \k 2
[e_=,{\tau}])
-\frac{\k}{4\p x}\int str [e_=,{\tau}] \bdel{\tau},\label{action0}
\eea
where
\bea
A&\in&\Pi_{>0}\cg ,\nonu
{\tau}&\in&\Pi_{\frac 1 2}\Pi_{\mbox{im ad}_{\dis e_\pp}}\cg
\eea
is gauge invariant:
\bea
\d g&=& \h g,\nonu
\d A&=& \bdel \h + {[}\h, A{]},\nonu
\d {\tau}&=& - \Pi_{\frac 1 2}\Pi_{\mbox{im ad}_{\dis
e_\pp}}\h,\label{trsf0}
\eea
and $\h\in \Pi_{>0}\cg$.
However, as we just discussed (and in the example in section 2), we have to
face the possibility
that constraints of the form \bea
\Pi_{<0} \Pi_{\mbox{ker ad}_{\dis e_\pp}} J =0, \eea
arise. In order to avoid this we introduce an extra term in the action
of the form
\bea
-\frac{\k}{2 \pi x}\int str\, A \Psi,\label{modif}
\eea
where $\Psi \in \Pi_{<0} \Pi_{\mbox{ker ad}_{\dis e_\pp}} \cg$. Of course the
action is now not invariant
anymore. Obviously one gets non-invariance terms of the form $\int str (
\delb\h\,\Psi)$. These can easily be
cancelled by adding a new field $\bar{\Psi}$, where $\bar{\Psi}
\in \Pi_{>0}
\Pi_{\mbox{ker ad}_{\dis e_=}} \cg$ which transforms as
\bea
\d \bar{\Psi}= \Pi_{\mbox{ker ad}_{\dis e_=}}\h. \label{trsf1}
\eea
We modify the action to $\cs_0+\cs_1$, where $\cs_0$
is given in eq. (\ref{action0}) and
\bea
\cs_1=-\frac{\k}{2\pi x}\int str\, A \Psi + \frac{\k}{2\pi x}\int str\, \Psi
\bdel \bar{\Psi} .
\eea
The resulting action is still not quite invariant. Indeed varying $\cs_0+\cs_1$
under
eqs. (\ref{trsf0}) and (\ref{trsf1}), yields
\bea
\d(\cs_0+\cs_1)=\frac{\k}{2 \pi x}\int str\, A [\h,\Psi].
\eea
We can further rewrite this using $\h=\P_{ker\,ad\, e_=}\h+ \P_{im\,ad\,
e_{\pp}}
\h$ and $A=\P_{ker\,ad\, e_=}A+ \P_{im\,ad\, e_{\pp}}
A$. The terms proportional to $\P_{ker\,ad\, e_=}A$ can be cancelled by
modifying the transformation rule of $\Psi$ while terms proportional to
$\P_{ker\,ad\,
e_=}\h$ are cancelled by adding extra terms proportional to $\bar{\Psi}$ to the
action. However, this
will leave us with, among others, a term proportional to
\bea
\int str \left( \Pi_{\mbox{im ad}_{\dis e_\pp}} A [
\Pi_{\mbox{im ad}_{\dis e_\pp}}\h,\Psi] \right),
\eea
which cannot be cancelled without the
introduction of new fields. Introducing new fields
would disrupt the  balance of degrees of freedom which could only be restored
by the introduction of a larger gauge symmetry.
We will get around this problem by modifying the definition of $\Psi$ which
will be allowed to have a component in the
image of $ad_{e_=}$. We will do this in such a way that $\Psi$ still has the
same number of degrees of freedom
as if it belonged exclusively to the kernel of $ad_{e_{\pp}}$ and such that the
highest weight gauge remains
non singular, {\it i.e.} $\Pi_{<0} \Pi_{\mbox{ker ad}_{\dis e_\pp}} J \neq 0$.
We split $\Pi_{>0}\cg$ in two parts: $\Pi_{>0}\cg = \cg_0\oplus\cg_1$ and
similarly for $\Pi_{<0}\cg$:
$\Pi_{<0}\cg = \bar{\cg}_0\oplus\bar{\cg}_1$ such that $\bar{\cg}_0 \perp
\cg_1$ and $\bar{\cg}_1 \perp
\cg_0$. The action
\bea
\cs_1=-\frac{\k}{2\pi x}\int str\, A \Psi + \frac{\k}{2\pi x}\int str\, \Psi
\bdel \bar{\Psi}- \frac{\k}{2\pi x}\int str\, A \,
[\bar{\Psi},\Psi] ,\label{actionew}
\eea
where $\Psi\in\bar{\cg}_0$ and $\bar{\Psi}\in\cg_0$ is invariant under
\bea
\d A&=&\bdel\h + [\h,A]\nonu
\d\bar{\Psi}&=&\P_{\cg_0}\left(\h+[\h,\bar{\Psi}]\right)\nonu
\d\Psi&=&\P_{\bar{\cg}_0}[\h,\Psi],\label{invgen}
\eea
provided the conditions
\bea
&&[\cg_0,\cg_0]\subseteq \cg_1\nonu
&&[\cg_1,\cg_1]\subseteq \cg_1\nonu
&&[\cg_0,[\cg_0,\cg_0]]\subseteq \cg_1\nonu
&&[\cg_0,[\cg_0,\cg_1]]\subseteq \cg_1,\label{allcon}
\eea
and similarly for $\bar{\cg}_0$ and $\bar{\cg}_1$,
are satisfied. The full gauge invariant action used for the reduction is then
simply $\cs=\cs_0+\cs_1$, where
$\cs_0$ is given in eq. (\ref{action0}) and $\cs_1$ in eq. (\ref{actionew}).
Rests us now to determine $\cg_0$
as a function of the choice of $\cg$ and the embedding.
\noindent 1. $osp(2|2)$ as a regular subalgebra of $osp(m|2n)$ or
$D(2,1,\a)$\footnote{Again we discard
the case where a sum of $osp(2|2)$'s is embedded. In that case we have to deal
with $sl(2|1)$ representations
which are not fully reducible, a pathology we want to avoid.}
\vspace{.3cm}
\noindent The adjoint representation of $\cg$ decomposes into $(0,0)$,
$(0,1/2)$ and $(0,1)$ representations
of $sl(2|1)$. The grading we consider is given by the action of $\frac 1 2
ad_{e_0}+ad_{u_0}$. It turns out
that the conditions (\ref{allcon}) are satisfied provided one chooses:
\bea
\cg_0&=&\Pi_{\mbox{ker ad}_{\dis e_=}}\Pi_{>0}\cg\nonu
\cg_1&=&\Pi_{\mbox{im ad}_{\dis e_\pp}}\Pi_{>0}\cg,\label{dada}
\eea
and similarly $\bar{\cg}_0= \Pi_{\mbox{ker ad}_{\dis e_\pp}} \Pi_{<0}\cg$ and
$\bar{\cg}_1 = \Pi_{\mbox{im ad}_{\dis e_=}} \Pi_{<0}\cg$. In order to show
this, it is sufficient to realize
that the elements of $\cg_0$ have charge $b=1/2$ and their grading is either
$1/2$ or $1$. This already
implies that $[\cg_0,\cg_0 ] =0$. So condition one, three and four are
satisfied. Furthermore, all strict
positively
graded elements of $\cg$ have either charge $b=1/2$ or $b=0$. Finally the only
charged elements with grades
1/2 or 1 belong neccessarily to $\cg_0$. From this the third condition follows.

\noindent 2. $sl(2|1)$ as a principal subalgebra of $sl(2j+1|2j)$ or
$sl(2j|2j+1)$.
\vspace{.3cm}
\noindent One can verify that already for $j=3/2$, the second of the conditions
(\ref{allcon}), gets
violated if one chooses $\cg_0=\Pi_{\mbox{ker ad}_{\dis e_=}}\Pi_{>0}\cg$.
However, if we choose
\bea
\cg_0=\Pi_{\mbox{ker ad}_{\dis e'_=}}\Pi_{\mbox{ker ad}_{\dis e_=}}\Pi_{>0}
\cg,\label{g0ch}
\eea
where $e'_=$ is $e_=$ restricted to the $sl(2j+1)$ subalgebra of $sl(2j+1|2j)$
or $sl(2j|2j+1)$, we find that
conditions (\ref{allcon}) are satisfied. Indeed, take first the case of
$sl(2j+1|2j)$ and parametrize it by
$(4j+1)\times (4j+1)$ matrices where the first $2j+1$ rows and columns are of a
bosonic nature, while the last
$2j$ rows and columns are fermionic. With this, one gets
\bea
e_0&=&\sum_{p=1}^{2j+1}2(j-p+1)E_{p,p} + \sum_{p=1}^{2j} 2 (j+\frac 1 2 - p)
E_{2j+1+p,2j+1+p}\nonu
e_=&=&\sum_{p=1}^{2j}E_{p+1,p}+ \sum_{p=1}^{2j-1}E_{2j+p+2,2j+p+1}\nonu
e'_=&=&\sum_{p=1}^{2j}E_{p+1,p}\nonu
u_0&=&-2j \sum_{p=1}^{2j+1}E_{p,p}-(2j+1)\sum_{p=1}^{2j}E_{2j+1+p,2j+1+p},
\eea
where $E_{r,s}$ is a $(4j+1)\times (4j+1)$ matrix unit: $(E_{r,s})_{kl} =
\d_{r,k}\d_{s,l}$. According to the
previous discussion, we choose the grading given by the action of $\frac 1 2
ad_{e_0}+2j\,ad_{u_0}$ and we
find that $\P_{>0}\cg$ is generated by $E_{k,l}$ with $k<l$. $\cg_0$ is
generated by $\{ E_{2j+1,2j+1+p}, p
\in \{1, 2, \cdots, 2j\}\}$, and we have {\it e.g.}
$\bar{\Psi}=\sum_{p=1}^{p=2j}\bar{\Psi}^p E_{2j+1,2j+1+p}$.
With this one verifies that the conditions (\ref{allcon}) are satisfied.
In particular, we have again that $[\cg_0 ,\cg_0]=0$. The reduction of the
adjoint representation is
\bea
\mbox{adjoint}\left(sl(2j+1|2j)\right)=\bigoplus_{p=1}^{2j} (0,p).\label{redad}
\eea
The discussion for $\cg=sl(2j|2j+1)$ is completely analogous. We represent
$\cg$ by $(4j+1)\times (4j+1)$
matrices of which the first $2j$ rows and columns are bosonic, while the last
$2j+1$ rows and columns are
fermionic. We have
\bea
e_0&=&\sum_{p=1}^{2j}2(j-p+\frac 1 2 )E_{p,p} + \sum_{p=1}^{2j+1} 2 (j - p + 1)
E_{2j+p,2j+p}\nonu
e_=&=&\sum_{p=1}^{2j-1}E_{p+1,p}+ \sum_{p=1}^{2j}E_{2j+p+1,2j+p}\nonu
e'_=&=&\sum_{p=1}^{2j}E_{2j+p+1,2j+p}\nonu
u_0&=&-(2j+1) \sum_{p=1}^{2j}E_{p,p}-2j\sum_{p=1}^{2j+1}E_{2j+p,2j+p}.
\eea
$\P_{>0}\cg$ is now generated by $E_{k,l}$ with $1\leq k<l\leq 2j$ or
with $2j+1\leq k<l\leq 4j+1$ or $2j+1\leq k\leq 4j+1$ and $1\leq l\leq 2j$.
$\cg_0$ is generated by
$\{ E_{4j+1,p}, p
\in \{1, 2, \cdots, 2j\}\}$ and the adjoint representation decomposes as in eq.
(\ref{redad}).

\noindent 3. $sl(2|1)$ as a principal subalgebra of $p\, sl(2j+1|2j) \oplus q
\, sl(2j|2j+1)$.
\vspace{.3cm}
\noindent This case follows immediately from the previous one. The grading is
still given by the action of
$\frac 1 2 ad_{e_0}+2j\,ad_{u_0}$ and $\cg_0$ is taken as in eq. (\ref{g0ch}),
but now $e'_=$ is $e_=$
restricted to the $p\, sl(2j+1) \oplus q \, sl(2j+1)$ subalgebra of $\cg$. The
adjoint of $\cg$ decomposes as
\bea
\mbox{adjoint}\left(p\, sl(2j+1|2j) \oplus q \, sl(2j|2j+1)\right)=
(p+q)\bigoplus_{r=1}^{2j} (0,r).\label{redad3}
\eea

\noindent 4. $sl(2|1)$ as a principal subalgebra of $p\, sl(2j+1|2j) \oplus q
\, sl(2j|2j+1)$, which in its
turn is regularly embedded in $ sl(p(2j+1)+2qj| 2pj +
q(2j+1))$.
\vspace{.3cm}
\noindent Again, we choose the action of $\frac 1 2 ad_{e_0}+2j\,ad_{u_0}$ as
the grading.
The choice of $\cg_0$ is exactly as in the previous case, {\it i.e.} one takes
eq. (\ref{g0ch}), with
$e'_=$ as $e_=$
restricted to the $p\, sl(2j+1) \oplus q \, sl(2j+1)$ subalgebra of the
regularly in $\cg$ embedded
$p\, sl(2j+1|2j) \oplus q \, sl(2j|2j+1)$ algebra.
To show this, one only has to show this to be true for the part of $\cg$ which
does not belong to the regularly embedded
$p\, sl(2j+1|2j) \oplus q \, sl(2j|2j+1)$ subalgebra. The $p\, sl(2j+1|2j)
\oplus q \, sl(2j|2j+1)$ subalgebra
follows from the previous. Those generators fall (using a slightly abusive
notation)
either in a $((\pm j,j),(\mp j,j))$ of an
$sl(2j+1|2j) \oplus sl(2j+1|2j)$ subalgebra, while being a scalar
for the other factors of $p\, sl(2j+1|2j) \oplus q \, sl(2j|2j+1)$, or in a
$((\pm j,j),(\mp j,j)^\p)$
representation of an $sl(2j+1|2j) \oplus sl(2j|2j+1)$ subalgebra. An explicit
parametrization of these two
subcases, as we did under case 2, quickly shows that the choice of $\cg_0$ is
consistent with eqs.
(\ref{allcon}). The adjoint representation decomposes as
\bea
&&\mbox{adjoint}\Big( sl(p(2j+1)+2qj| 2pj +
q(2j+1))\Big)=
(p+q)\bigoplus_{r=1}^{2j} (0,r)\nonu
&&\ \ \oplus \Big(p(p-1)+q(q-1)\Big)\bigoplus_{r=0}^{2j} (0,r)
\oplus 2pq\bigoplus_{r=0}^{2j} (0,r)'.\label{redad4}
\eea
\sect{Quantizing the model \label{qtm}}
Our starting point is the action $\cs_0+\cs_1$, where $\cs_0$ is given in eq.
(\ref{action0}) and $\cs_1$ in
eq. (\ref{actionew}). The gauge invariance eq. (\ref{invgen}) is fixed by the
choice $A=0$. We introduce
ghosts $c\in\P_{>0}\cg$ and anti-ghosts $b\in\P_{<0}\cg$ and obtain the gauge
fixed action:
\bea
{\cal S}_{\rm gf}=\k S^-[g]
+ \frac{\k}{4\p x}\int str [\t,e_=]\bdel\t+ \frac{\k}{2\pi x}\int str\, \Psi
\bdel \bar{\Psi}
+\frac{1}{2\p x}\int str\, b\bdel c,
\label{actiongf1}
\eea
with the BRST charge $\cq_{HR}$:
\bea
\cq_{HR}=\frac{1}{4\p i x}\oint str\left\{ c \left( J -\frac \k 2
e_=-\frac \k 2[e_=,\t]-\frac\k 2 \Psi- \frac{\k}{2}
[\bar{\Psi},\Psi] + \frac 1 2 J^{\rm gh}\right) \right\}.\label{brsch2}
\eea
where $J^{\rm gh}=\frac 1 2 \{b,c\}$. The conformal currents are obtained as
the generators of the cohomology
$H^*(\ca,\cq_{HR})$, where $\ca$ is the algebra generated by
$\{b,\hat{J}=J+J^{\rm
gh},\t,\Psi,\bar{\Psi},c\}$. Every field has a double grading $(m,n)$ where $m$
is the grading used in the
reduction and $m+n$ is the ghost number of the field. The fields $\t$, $\Psi$
and $\bar{\Psi}$ have grading
$(0,0)$. Again the BRST operator decomposes into pieces of definite grading. A
novel feature which appears
here is the fact that in general there will be more than 3 pieces! A similar
situation occured in the
construction of topological strings from Hamiltonian reduction \cite{LLS}. We
decompose $\cq_{HR}$ as
$\cq_{HR}=\cq^\Psi+\cq_{(1,0)} + \cq_{(1/2,1/2)} +\cq_{(0,1)}$ where $\cq^\Psi$
is the $\Psi$ dependent part
of $\cq_{HR}$ and the remainder of the decomposition is as in eq. (\ref{qsqr}).
If $\Psi$ has gradings
$-1/2,-1,\cdots,-n_{max}$ then $\cq^\Psi$ decomposes as
\bea
\cq^\Psi=\sum_{p=1}^{2n_{max}-1}\cq^\Psi_{(\frac 1 2+\frac p 2,\frac 1 2-\frac
p 2)}.
\eea
The nontrivial action of $\cq_{HR}$ on various fields is tabulated below:
\bea
\cq^\Psi:&&b\rightarrow -\frac\k 2
\left(\Psi+\P_{<0}\left[\bar{\Psi},\Psi\right]\right)\nonu
&&\Psi\rightarrow\frac 1 2 \P_{\bar{\cg}_0}[c,\Psi]\nonu
&&\bar{\Psi}\rightarrow\frac 1 2 \P_{\cg_0}(c+[c,\bar{\Psi}]\nonu
&&\hat{J}\rightarrow\frac\k 4 [c,\Psi+[\bar{\Psi},\Psi]]\nonu
\cq_{(1,0)}:&&b\rightarrow -\frac\k 2 e_=\nonu
&&\hat{J}\rightarrow -\frac\k 4 [e_=,c]\nonu
\cq_{(\frac 1 2 , \frac 1 2)}:&&b\rightarrow -\frac\k 2 [e_=,\t]\nonu
&&\t\rightarrow - \frac 1 2 \P_{\frac 1 2}\Pi_{\mbox{im ad}_{\dis
e_\pp}}c\nonu
&&\hat{J}\rightarrow -\frac\k 4 [[e_=,\t],c]\nonu
\cq_{(0,1)}:&&b\rightarrow\P_{<0}\hat{J}\nonu
&&\hat{J}\rightarrow
\frac 1 2 \vec{[}c,\P_{\geq 0}\hat{J}]+\frac \k
4 \del c -\frac 1 2 [\P_{<0}(t^A),[\P_{>0}(t_A),\del c ]]
\nonu
&&c\rightarrow\frac 1 2 cc,\label{Qtrsf}
\eea
where $\vec{[}A,B]$ stands for
\bea
\vec{[}X,Y]=(-)^{(AB)}\left(X^AY^B \right)f_{AB}{}^Ct_C,
\eea
with $\left(X^AY^B \right)$, a regularized product.
Using eq. (\ref{Qtrsf}), one shows that the cohomology $H^*(\ca,\cq_{HR})$ is
isomorphic to
$H^*(\wha,\cq_{HR})$, where $\wha$ is a reduced complex generated by
$\{\P_{\geq 0}\hat{J},\t,\Psi,\bar{\Psi},c\}$. The double grading introduced
before, carries over to the reduced complex:
\bea
\wha=\bigoplus_{p\in\frac 1 2{\bf N}}\bigoplus_{m\in{\bf N}}\wha_{(p,-p+m)}.
\eea
We introduce a filtration, which in the case of $n_{max}\leq 1$ is given by
\bea
\wha^m=\bigoplus_{k\in\frac 1 2 {\bf Z}}\bigoplus_{l\geq m}\wha_{(k,l)}
\eea
otherwise
\bea
\wha^m=\bigoplus_{k\in \frac 1 2 {\bf Z}}\bigoplus_{l\geq
\frac{1-n_{max}}{n_{max}}k+m}\wha_{(k,l)}.
\eea
The rest is now quite standard. One sets up a spectral sequence $(E_r, d_r)$,
$r\geq 1$,
which in the first case
collapses after two steps: $E_2=E_\infty$, and in the second case after
$2n_{max}$ steps: $E_{2n_{max}}
=E_\infty$. The actual computation is quite involved. As an example we
explicitely compute the sequence for
the case of a regular $osp(2|2)$ embedding. In that case one deals only with
$(0,0)$, $(0,1/2)$ and $(0,1)$
irreps of $sl(2|1)$. As explained, we have that the grading is given by the
action of
$\frac 1 2 ad_{e_0}+ad_{u_0}$, and the choice for $\cg_0$ and $\cg_1$ is given
in eq. (\ref{dada}). From this
one gets that $\cg_0$ has elements of grade $1/2$ and $1$ only and so it
follows that
$\cq_{HR}=\cq_{(1,0)} + \cq_{(1/2,1/2)} +\cq_{(0,1)}$, where we absorbed the
$\Psi$ dependent parts of
appropriate grading in  $\cq_{(1,0)}$ and  $\cq_{(1/2,1/2)}$. Using eq.
(\ref{Qtrsf}) one finds:
\bea
&&E_1=H^*(E_0;d_0)=H^*(\wha ; \cq_{(1,0)}) =\ca (\P_{\frac 1 2}c) \otimes
\ca ( \Psi) \otimes \ca (\P_{\frac 1 2}\bar{\Psi} ) \otimes\nonu
&&\qquad \ca (\t) \otimes \ca \left(\Pi_{\mbox{ker ad}_{\dis
e_\pp}}
\Pi_{\geq 0}\left(\hat{J}-\frac \k 2 [\bar{\Psi},\P_{-1}\Psi]+[L\circ\P_{\geq
0}\hat{J},
\P_{-1}\Psi]\right)\right),
\eea
where $L$ is defined by
\bea
L\circ \mbox{ad}_{\dis
e_=}=\Pi_{\mbox{im\, ad}_{\dis
e_\pp}},\qquad \mbox{ad}_{\dis
e_=}\circ L=\Pi_{\mbox{im\, ad}_{\dis
e_=}}.
\eea
Subsequently we get
\bea
&&E_2=E_\infty=H^*(E_1;d_1)=H^*(E_1 ; \cq_{(\frac 1 2 , \frac 1 2)}) =\nonu
&&\qquad \ca \left( \Pi_{\mbox{ker ad}_{\dis
e_\pp}}\Pi_{\geq 0}
\left(\hat{J}+\frac\k 4 [\t,[e_=,\t]] -\frac \k 2
[\P_{1}\bar{\Psi},\P_{-1}\Psi]-\frac \k 2 [\P_{\frac
1 2}\bar{\Psi},\P_{-\frac 1 2}\Psi]+\right.\right.\nonu
&&\qquad\left.\left. [L\circ\P_{\geq 0}\hat{J},
\P_{-1}\Psi]\right)\right)\otimes \ca \left(\Pi_{\mbox{ker ad}_{\dis
e_\pp}}\Pi_{<0}\left(\Psi +[\t,\Psi]\right)\right).
\eea
The full generators are then obtained by a generalized tic-tac-toe
construction. Finally for $(0,1/2)$ and $(0,0)$ multiplets, we introduce
scalars and fermions to complete the superconformal representations through a
reversed Goddard-Schwimmer mechanism as was explained in section 5.

However, one more point remains to be clarified. As advertised before, we would
like to identify the
generators (which are already BRST invariant) $\Pi_{\mbox{ker ad}_{\dis
e_\pp}}\Pi_{<0}(\Psi +[\t,\Psi]+\mbox{ contributions arising from the inverse
Goddard-Schwimmer mechanism})$ with the anti-ghosts. An
anti-ghost is a simple field, while the
previous expression contains composite terms albeit of a very simple nature.
This problem was solved in
\cite{BLLS} through the introduction
of a similarity transformation generated by $S=\exp \, R$ with
\bea
R&\propto& \frac{1}{2\pi i}\oint(\bar{\Psi}[\tau, \Psi]+
\mbox{ contributions arising from the}\nonu
&&\mbox{inverse
Goddard-Schwimmer mechanism}).
\eea
\sect{Discussion}
In this paper we classified all possible embeddings of $sl(2|1)$ into super Lie
algebras. This classification
is equivalent to the classification of all extended $N=2$ superconformal
algebras which can be obtained from
Hamiltonian reduction. While a specific embedding fixes the conformal algebra,
including the value of the
central charge, completely, the particular
realization of that algebra is only determined once a grading on the super Lie
algebra has been chosen. The
canonical grading, which is just the $sl(2)$ grading inherited from the
embedding, yields the standard or
symmetric realizations. Twisting or modifying the grading by adding a multiple
of the $u(1)$ charge to the
$sl(2)$ grading results in realizations which allow for a stringy
interpretation.
After the reduction we are left with the $N=2$ superconformal currents together
with a set of $N=2$ multiplets each of  which generically contains four
currents which yields some extension of the $N=2$ superconformal algebra. The
currents fall generically (we
will see that chiral and anti-chiral multplets do not have to be considered)
into unconstrained $N=2$
multiplets each containing four currents, say $Y(x)$, $H_+(x)$, $H_-(x)$ and
$Z(x)$ of conformal dimensions
$h+1$, $h+1/2$, $h+1/2$ and $h$. Twisting
amounts to replacing $Y(x)$ by $X(x)\equiv Y(x)+\frac 1 2 \del Z(x)$. The OPE's
of the twisted $N=2$ subalgebra
itself were given in eq. (\ref{twn2}). The OPE's of $T=T_{N=2}+\frac 1 2\del
U$,
$G_\pm$ and $U$ with $X(x)$, $H_+(x)$, $H_-(x)$ and
$Z(x)$ follow immediately from eq. (\ref{twn4}). We give the most significant
ones:
\bea
&&T(z_1)X(z_2)=(h+1-\frac q 2)z_{12}^{-2}X(z_2)+z_{12}^{-1}\del X(z_2),\nonu
&&T(z_1)H_-(z_2)=(h+1-\frac q 2)z_{12}^{-2}H_-(z_2)+z_{12}^{-1}\del
H_-(z_2),\nonu
&&G_+(z_1)H_-(z_2)= (h+\frac q 2)z_{12}^{-2}Z(z_2) +
z_{12}^{-1} X(z_2),\nonu
&&U(z_1)X(z_2)=(h+\frac q 2)z_{12}^{-2}Z(z_2)+q z_{12}^{-1}X (z_2),
\label{twn3}
\eea
If now, $G_-$ and all fields of the $H_-$ type are realized as single fields,
something which was achieved in this paper,
then we can view the above system as a string theory. Indeed $Q$,
\bea
Q=\frac{1}{2\p i}\oint dz G_+(z),
\eea
is the BRST charge, satisfying $Q^2=0$. The $G_-$ current and all currents of
the $H_-$ type are the anti-ghosts. The total symmetry currents (matter +
gravity + ghosts) are the energy-momentum tensor $T=T_{N=2}+\frac 1 2 \del U$
and the currents of the $X$ type. One notices that w.r.t. the twisted
energy-momentum tensor $X$ has become primary, this in contrast with the
situation of $Y$ vs. $T_{N=2}$. One verifies from eq. (\ref{twn3}) that
they are indeed the BRST transform of the corresponding antighosts:
\bea
T=[Q,G_-]_+\qquad\quad X=[Q,H_-]_\pm.
\eea
However, one more restriction follows. We saw that $U$ has, in leading order,
the interpretation of a ghost number current. So we should consider those
reductions where in eq. (\ref{twn3}) only $q=0$ appear. This is equivalent to
restricting to those $sl(2|1)$ embeddings where in the decomposition of the
adjoint representation, only $sl(2|1)$ irreps $(b,j)$ with $b=0$ occur. They
were classified in section 6.
The question which obviously arises is:
given an $sl(2|1)$ embedding,
which string theory do we describe? It is straightforward to see that for the
two main cases we obtain
the following pattern\footnote{Here and elsewhere, we
use the symbol $\rightarrow$ for a principal embedding, while $\hookrightarrow$
for a regular embedding.}:\\
I. $sl(2|1)\rightarrow p\,sl(2j+1|2j)\oplus q\, sl(2j|2j+1)\hookrightarrow
sl(p(2j+1)+2qj|2pj+q(2j+1))$\\
The matter sector of the string theory corresponds with the reduction:
\bea
sl(2)\rightarrow p\, sl(2j+1)\oplus q\, sl(2j+1)\hookrightarrow
sl(p(2j+1)|q(2j+1)).
\eea
II. $osp(2|2)\hookrightarrow osp(m|2n)$\\
The matter sector is now given by the reduction
\bea
sp(2)\hookrightarrow sp(2n) \hookrightarrow osp(m-2|2n).
\eea
It remains a very interesting open question which string theories arise from
the reduction of $D(2,1,\a)$. One would expect $N=2$ strings. However in
\cite{BLLS} it was shown that the standard $N=2$ strings arise from the
reduction of $osp(4|2)$ which turns out to be isomorphic to $ D(2,1,\a)$ with
$\a =1$. Presumably, one will get a new type of $N=2$ string. In view of the
very particular properties of $N=2$ strings \cite{oog}, this case definitely
needs further investigation. Work in this direction is in progress.

Explicitely worked out examples of the general method developed in this paper
can be found in the literature.  In \cite{BLNW} e.g., classical $W_n$ strings
were obtained from a reduction based on $sl(2|1)\rightarrow sl(n|n-1)$. The
quantum structure can now also be
obtained following the strategy developed in previous section. In \cite{BLLS},
$N$-extended superstrings were
obtained from the reduction $osp(2|2)\hookrightarrow osp(N+2|2)$.
One can now wonder whether, in the case of an embedding where the adjoint
representation decomposes into $sl(2|1)$ irreps $(b,j)$, where $b$ is not
neccessarily zero, some stringy interpretation can still be given.  Similarly,
another point of
interest is the occurence non-fully
reducible representations in the decomposition of the adjoint representations
for certain $osp(2|2)$ embedings. An immediate consequence of this is that
there must exist non-fully reducible  $N=2$
superconformal representations. One might wonder whether such representations
might provide clues to the open problem of finding an off-shell description of
certain $N=2$ non-linear $\sigma$-models \cite{martin}. These
questions are presently under study and the results will be reported on
elsewhere.
Finally, a most interesting point would be to push the present work further and
address questions such as ``What is the spectrum of these stringtheories?''.
Using the recent results in \cite{driel} it should be possible to obtain at
least the partition function
explicitely.

\vspace{1cm}

\noindent{\bf{Acknowledgements:}} Two of us (E.R. and P.S.) wish to thank M.
Scheunert for useful informations about $sl(2|1)$ irreps and L. Frappat for
fruitful discussions.
\setcounter{section}{0}
\startappendix
\section{Wess-Zumino-Witten Models}
\setcounter{footnote}{0}
We briefly review WZW models.
Given a super
Lie algebra with generators $\{ t_a ; a\in\{ 1,\cdots,d_B+d_F\}\}$, where
$d_B$ ($d_F$) is the number of bosonic (fermionic) generators, we denote the
(anti)commutation relations by
\be
[t_a, t_b]=t_at_b-(-)^{(a)(b)}t_b t_a=f_{ab}{}^c t_c, \ee
where for $t_a$, $(a)=0\ (1)$ when $t_a$ is bosonic (fermionic). We always use
that
$X t_a= (-)^{(X)(a)} t_a X $ where $X$ is not Lie algebra valued. {}The adjoint
representation is \be
[t_a]_b{}^c\equiv f_{ba}{}^c.
\ee
The Killing metric $g_{ab}$ is defined by \be
f_{ca}{}^d
f_{db}{}^c(-)^{(c)} = - \tilde{h} g_{ab}, \ee
with $\tilde{h}$, the dual Coxeter
number. This is fine for Lie algebras, but for super algebras the dual Coxeter
number might vanish. More generally we have then \be
str(t_at_b)\equiv [t_a]_{\a}{}^{\b}[t_b]_{\b}{}^{\a}(-)^{(\a )}\equiv -x g_{ab}
\ee
where $x$ is the index of the representation. In the adjoint representation one
has $x=\tilde{h}$. A
contraction runs from upper left to lower
right, {\it e.g.} $A^aB_a$. Raising and lowering indices happen according to
this convention
(implying $g^{ac}g_{bc}=\d^a_b$):
\be
A^a=g^{ab}A_b\qquad A_a=A^bg_{ba}.
\ee
We tabulate some properties of the (super) Lie algebras which appear in this
paper:
\begin{center}
\begin{tabular}{||c||c|c|c|c||} \hline
& & & &  \\
algebra & bosonic & {$d_B$} & {$d_F$} & {$\tilde{h}$}  \\
& subalgebra& && \\ \hline
{$osp(m|2n)$}&{$so(n)\oplus sp(2n)$}& {$ {\scriptstyle \frac 1 2 m(m-1) +
n(2n+1) }$}&{$ {\scriptstyle
2mn }$} &
{${\scriptstyle \frac 1 2 m-n-1 }$} \\ \hline
{$D(2,1,\a)$}&{$sl(2)\oplus sl(2)\oplus sl(2)$}&{${\scriptstyle 9}$} &{$
{\scriptstyle 8}
$} & {$ {\scriptstyle 0}$}  \\ \hline {$sl(m|m)$}&{$sl(m)
\oplus sl(m)$}&
{${\scriptstyle 2 (m^2-1)}$} &{$ {\scriptstyle 2 m^2} $} & {$
{\scriptstyle 0}$}  \\ \hline
{$sl(m|n)$}&{$sl(m)+sl(n)+gl(1)$}&
{${\scriptstyle m^2+n^2-1}$} &{$ {\scriptstyle
2mn} $} & {$ {\scriptstyle m-n}$} \\ {$m\neq
n$}& & & & \\ \hline
{$g(3)$}&{$g_2\oplus sp(2)$}&{${\scriptstyle 17}$} &{$ {\scriptstyle 14}
$} & {$ {\scriptstyle \frac 1 4}$}  \\ \hline
{$f(4)$}&{$so(7)\oplus sp(2)$}&{${\scriptstyle 21}$} &{$ {\scriptstyle 16}
$} & {$ {\scriptstyle \frac 1 2}$}  \\ \hline
\end{tabular}
\end{center}
The WZW action $\k S^+[g]$ is given by
\be
\k S^+ [g]= \frac{\k}{4\p x} \int d^2 z \; str \left\{ \del g^{-1} \bar{\del} g
\right\} + \frac{\k}{12\p x}
\int d^3 z\; \e^{\a\b\g} \, str \left\{ g_{,\a} g^{-1} g_{,\b}
g^{-1} g_{,\g} g^{-1} \right\},
\label{nineteen}
\ee
and  satisfies the Polyakov-Wiegman identity, \be
S^+[hg]=S^+[h]+S^+[g]-\frac{1}{2\pi x}\int str\Bigl( h^{-1}\del h \bdel g
g^{-1} \Bigr) .\label{pwfor}
\ee
The functional $S^-[g]$ is defined by
\be
S^-[g]=S^+[g^{-1}].
\ee
Using the equations of motion
\bea
\d S^+[g]&=&\frac{1}{2\p x}\int str\left\{\bdel(g^{-1}\del g)g^{-1}\d
g\right\}\nonu
&=&\frac{1}{2\p x}\int str\left\{\del(\bdel g g^{-1})\d g g^{-1}\right\},
\eea
which are solved by putting $g\equiv g(\bz )g(z)$ where $\del g(\bz )=\bdel g(z
)=0$, one gets
the conserved affine currents
\bea
J_z&=&-\frac{\k}{2}g^{-1}\del g \nonu
J_{\bz}&=&\frac{\k}{2}\bdel g g^{-1}
\eea
which generate the affine symmetries \bea
\d J_z^a&=&-\frac{\k}{2} \del\h^a-
(-)^{(b)(c)}f_{bc}{}^{a}\h^bJ_z^c\nonu
\d J_{\bz}^a&=& \frac{\k}{2}\bdel\bar{\h}^a+ (-)^{(b)(c)}f_{bc}{}^{a}
\bar{\h}^b
J_{\bz}^c
\eea
where
\be
\bdel\h^a=\del\bar{\h}^a=0.
\ee
{}From
\be
\d J_z^a(z)=\frac{1}{2\p i}\oint_z dw J_z^b(w)\h_b(w)J^a_z(z), \ee
we get the OPE of an affine Lie algebra of level $\k$: \be
J^a_{z} (z) J^b_{z} (w) = - \frac{\k}{2} g^{ab} (z-w)^{-2} + (z-w)^{-1}
(-)^{(c)}f^{ab}{}_c
J^c_{z} (w) + \cdots,
\label{ven}
\ee
and similarly for $J_{\bz}$. The Sugawara construction for the energy-momentum
tensor is given by \be
T=\frac{1}{x\left(\k+\tilde{h} \right)}str J_zJ_z, \ee
and it satisfies the Virasoro algebra with the central extension given by:
\be
c=\frac{\k (d_B-d_F)}{\k+\tilde{h}}.
\ee

\end{document}